\documentclass[acmsmall]{acmart}

\setcopyright{rightsretained}
\acmJournal{PACMMOD} \acmYear{2024} \acmVolume{2} \acmNumber{1 (SIGMOD)}
\acmArticle{58} \acmMonth{2} \acmPrice{}\acmDOI{10.1145/3639313}

\usepackage{algorithm}
\usepackage{algpseudocode}
\usepackage{color, soul}
\usepackage{xcolor}
\usepackage[frozencache,cachedir=.]{minted}
\usepackage{array}
\usepackage{listings}
\usepackage{tikz}
\usepackage{caption}
\usepackage{subcaption}
\usetikzlibrary{positioning}
\usetikzlibrary{calc,bending}
\usepackage{mathpartir}
\usepackage{float}
\usepackage{stfloats}
\usepackage{enumitem}
\usepackage{float}
\usepackage{marginalia}

\colorlet{light-gray}{gray!10}

\definecolor{highlight-green}{rgb}{0.6, 0.81, 0.66}
\definecolor{border-blue}{rgb}{0.38, 0.25, 0.34}

\definecolor{codegray}{gray}{0.9}
\newcommand{\code}[1]{\texttt{#1}}

\def\system{\textsc{Dias}}

\soulregister{\system}{0}

\definecolor{ddkang-blue}{rgb}{0.69, 0.87, 0.94}

\definecolor{stef-color}{rgb}{1, 0.898, 0.706}
\DeclareRobustCommand{\stef}[1]{}

\definecolor{charith-color}{rgb}{1, 0.5, 0.5}

\definecolor{revis-color}{rgb}{0.1, 0.1, 1}
\DeclareRobustCommand{\revis}[1]{{#1}}

\setminted{fontsize=\small} 

\newcommand{\minihead}[1]{\paragraph{\textbf{#1.}}}

\makeatletter
\gdef\@copyrightpermission{
  \begin{minipage}{0.2\columnwidth}
   \href{https://creativecommons.org/licenses/by/4.0/}{%
     \includegraphics[width=0.90\textwidth]{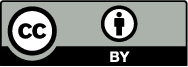}}
  \end{minipage}\hfill
  \begin{minipage}{0.8\columnwidth}
   \href{https://creativecommons.org/licenses/by/4.0/}{This work is licensed under a Creative Commons Attribution International 4.0 License.}
  \end{minipage}
  \vspace{5pt}
}
\makeatother

\begin{document}


\title[Dias: Dynamic Rewriting of Pandas Code]{%
\system{}: Dynamic Rewriting of Pandas Code}

\author{Stefanos Baziotis}
\affiliation{%
  \institution{University of Illinois (UIUC)}
  \city{Champaign-Urbana}
  \country{U.S.A}
}
\email{sb54@illinois.edu}

\author{Daniel Kang}
\affiliation{%
  \institution{University of Illinois (UIUC)}
  \city{Champaign-Urbana}
  \country{U.S.A}
}
\email{ddkang@illinois.edu}

\author{Charith Mendis}
\affiliation{%
  \institution{University of Illinois (UIUC)}
  \city{Champaign-Urbana}
  \country{U.S.A}
}
\email{charithm@illinois.edu}

\begin{abstract}

In recent years, dataframe libraries, such as \texttt{pandas} have exploded in
popularity. Due to their flexibility, they are increasingly used in
\emph{ad-hoc} exploratory data analysis (EDA) workloads. These workloads are
diverse, including custom functions which can span libraries or be written in
pure Python. The majority of systems available to accelerate EDA workloads focus
on bulk-parallel workloads, which contain vastly different computational
patterns, typically within a single library. As a result, they can introduce
excessive overheads for ad-hoc EDA workloads due to their expensive optimization
techniques. Instead, we identify source-to-source, external program rewriting as
a lightweight technique which can optimize across representations, and offer
substantial speedups while also avoiding slowdowns. We implemented \system{},
which rewrites notebook cells to be more efficient for ad-hoc EDA workloads. We
develop techniques for efficient rewrites in \system{}, including checking the
preconditions under which rewrites are correct, dynamically, at fine-grained
program points. We show that \system{} can rewrite individual cells to be
57$\times$ faster compared to \code{pandas} and 1909$\times$ faster compared to
optimized systems such as \code{modin}. Furthermore, \system{} can accelerate
whole notebooks by up to 3.6$\times$ compared to \code{pandas} and 27.1$\times$
compared to \code{modin}.

\end{abstract}



\received{July 2023}
\received[revised]{October 2023}
\received[accepted]{November 2023}

\maketitle

\section{Introduction}
\label{sec:Intro}

In recent years, \emph{dataframe}-based libraries, such as \texttt{pandas}, have
become increasingly popular with users ranging from social scientists to
business analysts \cite{pandas_in_business, pandas_in_social}. This growth is driven by many reasons, including
the flexibility of such libraries, the ability to work within a notebook
environment, and the interoperability with other libraries.

Due to the popularity of dataframe libraries, academic and industrial work has
focused on improving the scalability of \texttt{pandas} \emph{in the context of
bulk-parallel operations}. For example, libraries including
\code{modin} \cite{modin}, \code{dask} \cite{dask}, and PySpark \cite{pyspark_web} focus on
parallel or distributed dataframes. Many of these libraries focus on scaling out
\texttt{pandas} across multiple servers, as \texttt{pandas} will fail if the
dataframe does not fit in main memory.

\begin{figure}
\begin{minted}[bgcolor=light-gray]{python}
for i in range(1, 15):
  lhs = DF_PH.loc[i, 'VendorID']
  rhs = DF_PH.loc[i-1, 'VendorID']
  if lhs == rhs:
    counter += 1
  else:
    counter = 1
  DF_PH.loc[i, 'discourse_nr'] = counter
\end{minted}
    \caption{Loop which accesses individual elements (source: Kaggle
    \cite{real_nb_for_loop}). This loop can be hundreds of times slower in
    bulk-parallel frameworks like \code{modin}, \code{dask} etc. and PolaRS,
    which are not optimized for individual accesses.}
    \label{fig:for_loop}
\end{figure}

However, there has been an emerging class of important workloads that operate on
a \emph{single} machine, combined with ad-hoc functions. For example, in our
conversations with law professors at Stanford University, we have found that
provisioning and managing distributed clusters is challenging and time-consuming
for social scientists. As a result, much of the work done by such social
scientists is done on a single machine. Similarly, Kaggle and Google Colab
provide single-machine notebooks for data scientists to explore datasets.
Furthermore, many tasks require custom user-defined functions (UDFs) that are
not well suited to working directly within the \texttt{pandas} API.

While these bulk-parallel dataframe libraries improve the horizontal scalability
of dataframes, as we show in this work, they unfortunately can \emph{fail to
accelerate a wide range of single-machine, ad-hoc workloads}. For example,
\code{modin}, \code{dask}, and PySpark are all 2-200$\times$ slower than \texttt{pandas} for a
range of operations: when interfacing with \texttt{numpy}, looping over
individual rows, and even for simple operations like multiplying two columns (on
a single machine). For example, a simple loop (Figure~\ref{fig:for_loop}) can be
many \textit{hundreds} of times slower (see Section~\ref{sub-sec:cmp-libs}).
Furthermore, all distributed dataframe libraries we are aware of do not maintain
full \texttt{pandas} compatibility, requiring domain experts to learn new
libraries.

We propose an alternative approach to address the \emph{vertical} scalability of
dataframe libraries: \emph{rewriting} notebook cells to accelerate dataframe
computations by utilizing faster, but semantically equivalent code sequences. To
understand the potential for rewriting notebook cells, consider the two cells in
Figure~\ref{fig:apply_only_math}. The first cell is a simplified cell from a
real-world, Kaggle notebook. The second cell is an optimized cell with identical
semantics. While identical semantically, the second cell can run up to
1000$\times$ faster, showing that simple rewrites of notebook cells can
accelerate workloads.

A natural question that emerges is why can't the users write optimized code
themselves. As witnessed in compilers, automatic tools that accelerate code can
reduce developer effort and improve comprehensibility. Furthermore,
several rewrite rules in this paper were \emph{not apparent to the
authors} (e.g., the rules used to perform the rewrites in
Figure~\ref{fig:concat-with-lists} and Figure~\ref{fig:split}), even after
devoting considerable time studying the internals of Python and \code{pandas}.
To understand how this translates to non-experts, there are whole videos and
articles dedicated to patterns for speeding up \code{pandas} code via manual
rewriting \cite{pygotham_apply_vectorized, pandas_opt_article_docs,
pandas_opt_video_heisler, pandas_opt_video_trabelsi, pandas_opt_article_dan,
pandas_opt_article_numba}. Even then, optimizing code correctly is challenging
for non-experts and can lead to subtle bugs.

To realize the vision of rewriting notebook cells transparently, we
propose \system{}, a tool that automatically rewrites notebook cells. As we
show, \system{} can accelerate notebook cells by up to 57$\times$ completely
transparently to the user. 

\begin{figure}
  \begin{subfigure}[t]{0.45\linewidth}
\begin{minted}[bgcolor=light-gray]{python}
def weighted_rating(x, m=m, C=C):
  v = x['vote_count']
  R = x['vote_average']
  return (v/(v+m) * R) + (m/(m+v) * C)


df.apply(weighted_rating, axis=1)
\end{minted}
    \caption{Loop through rows (extracted from a Kaggle notebook
    \cite{real_nb_apply_math}). This, effectively, loops sequentially over each
    row, and the operations are performed in the Python interpreter.}
    \label{fig:apply_only_math_orig}
  \end{subfigure}
  \hfill
  \begin{subfigure}[t]{0.45\linewidth}
\begin{minted}[bgcolor=light-gray]{python}
def weighted_rating(x, m=m, C=C):
  v = x['vote_count']
  R = x['vote_average']
  return (v/(v+m) * R) + (m/(m+v) * C)

# Pass the whole `df` directly.
weighted_rating(df)
\end{minted}
    \caption{The function contains only column operations and thus can be applied directly to the whole \code{DataFrame}.}
    \label{fig:apply_only_math_rewr}
  \end{subfigure}
  \caption{A rewrite example where we avoid \code{apply()}. The rewritten
  version, which uses vectorized, native execution, can run up to 1000$\times$ faster.}
  \label{fig:apply_only_math}
\end{figure}

\system{} is the first source-to-source, dynamic rewriter for a Python library.
Source-to-source rewriting/compilation, especially over Python code, is
challenging because source languages are high-level, while traditional compiler
optimizations work better in low-level intermediate representations. But, we
observe that operating at the source level creates novel opportunities. One such
opportunity is that it enables us to rewrite across different representations
(e.g., Python and \code{pandas}). Such rewrites are high-level (leading to
significant speedups), and thus, the source language is a suitable
representation for performing them. It may be useful to compare and contrast a
high-level rewrite like the one in Figure~\ref{fig:concat-with-lists} with a
traditional low-level rewrite such as \code{a*2} $\rightarrow$ \code{a<<1}.

To perform these rewrites, \system{} needs to operate externally, in contrast with
previous work (e.g., TensorFlow Grappler \cite{tf_grappler}, \code{modin}
\cite{modin}, BELE \cite{best_effort_lazy}), which are implemented as libraries.
Namely, it views all user's code compared to just library code and can modify
any part of it. This allows \system{} to rewrite across library boundaries (e.g.,
Figure~\ref{fig:concat-with-lists}), which is infeasible with current
techniques. 

However, implementing a source-to-source, external rewriter introduces new
challenges. First, \system{} needs to understand more than one representation (i.e.,
\code{pandas} and Python). Second, providing any guarantees when operating in a
high-level language is naturally harder, especially when it comes to Python
which has no formal semantics and liberal typing and scoping rules. A specific
challenge was performing the precondition checks at the correct program points,
as we explain in Section~\ref{sub-sec:rewriter}. Finally, \system{} must operate
within interactive time scales: the overhead of rewriting cannot dominate cost
savings.


We designed \system{}' rewrite engine with two components: a pattern matcher
and a rewriter with design decisions that specifically address the
aforementioned challenges. The rewrite engine is lightweight, with a fast
pattern matcher that can quickly match patterns that we can rewrite into faster
versions and a rewriter which emits, or performs, necessary static and runtime
precondition checks to guarantee correctness, within interactive latencies.


We show that on real-world Kaggle notebooks, \system{} can accelerate cells by
up to 57$\times$ (1.27$\times$ geometric mean) and whole notebooks by up to
3.6$\times$ (1.31$\times$ geometric mean). We also compare \system{} with
\code{modin} and show that it can be up to 27.1$\times$ faster for whole
notebooks (4.9$\times$ geometric mean). Furthermore, \system{} can avoid
rewriting cells that cause slowdowns, resulting in overheads that are only due
to the pattern matcher. We show that these overheads, even in the cases where
cells are not rewritten, are below noise thresholds. Finally, \system{} uses
no extra memory or disk capacity.

In summary, we make the following contributions.

\begin{enumerate}
    \item We identify program rewriting as a lightweight technique to speed up
    \code{pandas}-heavy EDA workloads.
    \item We develop \system{}, the first external rewriter to rewrite code
    across different representations in a dynamic setting. We introduce rewrite
    rules that can significantly speed up \code{pandas} code, including ones
    that cross library boundaries.
    \item \system{} applies these rewrite rules automatically, at runtime, and
    verifies whether applying a rule is correct by injecting checks at
    fine-grained program points.
    \item We evaluate \system{} on real-world notebooks and show that it can
    speed up cells by up to 57$\times$ and notebooks by up to 3.6$\times$, with
    almost no memory or disk overheads. We further compare \system{}
    with \code{modin} \cite{modin} and show that it can be up to 27.1$\times$
    faster for whole notebooks (4.9$\times$ geometric mean).
\end{enumerate}

\section{Background}
\label{sec:Background}

\subsection{Setting}

In this work, we focus on the broad class of workloads commonly referred to as
\emph{exploratory data analytics} (EDA) \cite{eda_notebooks}. In EDA workloads,
the data is iteratively analyzed for interesting patterns. Since the patterns of
interest are unknown ahead of time, much of this work is done interactively, in
a notebook environment (e.g., Jupyter notebooks, other REPLs) using a dataframe
library. We focus on \texttt{pandas} and related libraries in this work.

In one common setting, analysts are interested in analyzing large datasets,
which typically do not fit in main memory on a single server. To accelerate such
workloads, both industry and academia have invested in accelerating
\emph{bulk-parallel} workloads.

Frameworks including \code{modin} \cite{modin} and \texttt{dask} \cite{dask} aim to
accelerate such workloads. They operate by providing APIs close to the standard
\texttt{pandas} API, distributing data across servers, and evaluating
functions lazily. When working within these libraries, they can accelerate workloads by
up to 100$\times$ \cite{modin}.

Unfortunately, these bulk-parallel libraries have several drawbacks. In
particular, these libraries were not designed for \emph{ad-hoc, single-machine}
workloads.

\minihead{Ad-hoc operations}
The primary drawback of these libraries is that they have poor support for
ad-hoc operations outside of the library API. For example, operations such as
looping over rows, column-wise operations that require intermediate
materialization for inspection (e.g., comparing a column to a constant), or
inspecting the first $n$ rows can be 30-1900$\times$ \emph{slower} than standard
\texttt{pandas} on a single machine. For example, as we explain in
Section~\ref{sub-sec:cmp-libs}, a simple loop, shown in
Figure~\ref{fig:for_loop}, can be many \textit{hundreds} of times slower.

\minihead{Single-machine overheads}
In addition to slowdowns for ad-hoc operations, these libraries can add
substantial memory overheads. We selected 20 random EDA notebooks from Kaggle
(under criteria described in Section~\ref{sub-sec:exp-setup}), which had
heavy \texttt{pandas} usage. \code{modin} generally increased memory usage, with
the peak memory usage being up to 772$\times$ higher than native
\texttt{pandas}. The peak memory usage increased 53$\times$ on average
(geometric mean).

\minihead{Usability}
In discussions with social scientists and law professors at Stanford University
and the University of California, Berkeley, we have found that learning new APIs
is challenging and time-consuming. In particular, these bulk-parallel libraries
are not direct drop-in replacements. To show this, we sampled 20 notebooks from
Kaggle at random (under criteria described in Section~\ref{sub-sec:exp-setup}).
Six of these notebooks (30\%) were unable to run when \texttt{pandas} was
replaced with \code{modin}.

Furthermore, setting up distributed clusters can be difficult in these settings.
As a result, the distributed speedups are difficult to realize in the settings
we focus on.

\hfill{}

In this paper, we introduce program rewriting as an automatic optimization
technique for Python code that interfaces with \texttt{pandas}, focusing on
accelerating single-server, ad-hoc EDA workloads.

\subsection{Rewriting as an alternative optimization}

Rewriting, for optimization purposes, is the process of replacing some part of code with a functionally equivalent but faster version.
Rewriting avoids the previously mentioned drawbacks of library-based optimization systems.
First, it inherently does not suffer from a lack of API support because it is not a replacement
for \code{pandas} and it can 
leave the code untouched if it cannot handle it.
Second, rewriting is a lightweight technique incurring minimal overheads, which 
scale proportionally only to the code, not the data.

\begin{figure}
  \begin{subfigure}{\columnwidth}
\begin{minted}[bgcolor=light-gray]{python}
pd.Series(df['A'].tolist() + df['B'].tolist())
\end{minted}
    \caption{Original: Concatenate \code{Series} by first turning them into
    lists. Extracted from a Kaggle notebook \cite{real_nb_concat}.}
    \label{fig:concat_orig}
  \end{subfigure}
  \hfill
  \begin{subfigure}{\columnwidth}
\begin{minted}[bgcolor=light-gray]{python}
pd.concat([df['A'], df['B']], ignore_index=True)
\end{minted}
    \caption{Rewritten: Use a \code{pandas}-provided function for concatenation}
    \label{fig:concat_rewr}
  \end{subfigure}
  \caption{Rewrite example that crosses library boundaries, and thus cannot be
  performed by previous techniques. The rewritten version can be up to 11$\times$
  faster.}
  \label{fig:concat-with-lists}
\end{figure}

Additionally, there are fundamental advantages \system{} has over library-based
optimization approaches. The rewrite system is transparent. When the user
observes a speedup, they can always see the code that the rewriter used. In
other words, the user does not need to understand the system to understand the
cause of the speedup. At the same time, the user's code remains intact. Further,
rewriting has the benefit of being able to optimize across library boundaries.
For example, \system{} can automatically perform the rewrite in
Figure~\ref{fig:concat-with-lists} (taken from a real-world notebook). The
original code crosses the library boundaries (twice!) as we move from
\code{pandas} to Python (by converting to a list) and then back to
\code{pandas}. To perform this rewrite, a tool needs to view all the code and
understand semantic equivalences and differences across library boundaries
(e.g., \code{pandas} and the host language, Python). This is not possible with
optimization approaches that purely aim at accelerating the \code{pandas} API.

\begin{figure}
  \begin{subfigure}{\columnwidth}
\begin{minted}[bgcolor=light-gray]{python}
df[['a', 'b']] = df['C'].str.split('(', expand=True)
\end{minted}
    \caption{Splitting a \code{pandas.Series} using
    \code{pandas.Series.str.split()}. Extracted from a Kaggle notebook \cite{real_nb_split}.}
    \label{fig:split_pandas}
  \end{subfigure}
  \hfill
  \begin{subfigure}{\columnwidth}
\begin{minted}[bgcolor=light-gray]{python}
a = []
b = []
ls = df['C'].tolist()
for it in ls:
    spl = it.split('(', 1)
    a.append(spl[0])
    b.append(spl[1] if len(spl) > 1 else None)
df['a'] = pd.Series(a, df['C'].index)
df['b'] = pd.Series(b, df['C'].index)
\end{minted}
    \caption{Splitting a \code{pandas.Series} in pure Python}
    \label{fig:split_python}
  \end{subfigure}
  \caption{Splitting in \code{pandas} and Python. Surprisingly, the pure Python
  implementation is up to 7$\times$ faster.}
  \label{fig:split}
\end{figure}

Rewriting appears simple, but it can be challenging when performed manually. There are many non-obvious rewrites that the user may not be able to discover easily. For example, it might seem that the only way to make \code{pandas} code faster through rewriting is by replacing it with other \code{pandas} code, or using a similar library such as \code{numpy}. This has been reinforced over years of data scientists being trained to remain within \code{pandas}/\code{numpy} as much as possible because these use native, vectorized implementations and are thus deemed to be much faster than pure Python. It might, then, be surprising that moving out of \code{pandas} and into pure Python can lead to significant speedups. One example is shown in Figure \ref{fig:split}. The task here is to split a \code{Series} of strings by the delimiter \code{'('}. The code in Figure~\ref{fig:split_pandas} (extracted from a Kaggle notebook) does it by using a \code{pandas}-provided function. One would expect that this is the best way to perform this operation. Nevertheless, the version in Figure \ref{fig:split_python} is 3.5$\times$ faster. 
It moves from \code{pandas} to pure Python (by converting \code{df['C']} to a Python list) and performs the operation with a sequential Python loop 
(in our case studies in Section~\ref{sub-sec:ablation}, we explain why this version is faster). 

It is unreasonable to expect general \code{pandas} users to comprehend Python, \code{pandas}, and \code{numpy} to such an extensive level to be able to discover such equivalent versions and evaluate their relative performance. Second, even if the user succeeds in these tasks, the rewritten version can be significantly harder to write and read, as is evident from Figure \ref{fig:split}. This can further lead to correctness concerns about the rewrite. Third, manual rewriting breaks the library abstraction. In the original code of Figure \ref{fig:split}, the user has to think only of \emph{what} \code{split()} does. But, to come up with the rewritten version, this abstraction's veil has to be removed as the user needs to think of \emph{how} to implement it.

These issues motivated us to build \system{}, a system that performs such rewrites \textit{automatically}, by guaranteeing \textit{correctness} and with minimal overhead. Section~\ref{sec:System} provides an overview of \system{}.



\section{\system{} Overview}
\label{sec:System}

\begin{figure}
\begin{tikzpicture}[
  node distance=2cm,
  font=\small,
  CodeBlock/.style = {
  align=center,
  rectangle,
  draw=border-blue,
  thick,
  },
  ComponentBlock/.style={
  rectangle,
  draw=blue,
  thick,
  fill=blue!20,
  text width=5em,
  align=center,
  rounded corners,
  minimum height=2em
  }
]

\node (codeSource) [CodeBlock] {%
\centering

\begin{minipage}{7.2cm}
{\fontfamily{qbk}\selectfont
Jupyter Cell Source
}
\centering
\begin{minted}[bgcolor=light-gray]{python}
print(...)
foo().sort_values().head(n=5)
\end{minted}
\end{minipage}%
};

\draw node[CodeBlock, below=0.7cm of codeSource] (PattMatcher) {%
\centering
\begin{minipage}{7.2cm}
\centering
{\fontfamily{qbk}\selectfont
Pattern Matcher
}

\begin{minted}[bgcolor=light-gray, escapeinside=||]{python}
print(...)
|\colorbox{highlight-green}{foo().sort\_values().head(n=5)}|
\end{minted}
\end{minipage}%
};

\draw node[CodeBlock, below=0.5cm of PattMatcher] (OnlRewr) {%
\centering
\begin{minipage}{7.2cm}
\centering
{\fontfamily{qbk}\selectfont
Rewrite the Code
}

\begin{minted}[bgcolor=light-gray, escapeinside=||]{python}
print(...)

def sort_head(tmp):
  if type(tmp) == pd.Series:
    return tmp.nsmallest()
  else:
    tmp.sort_values().head(n=5)

sort_head(foo())

\end{minted}
\end{minipage}%
};

\draw node[CodeBlock, below=0.7cm of OnlRewr] (RunCell) {%
\centering
\begin{minipage}{7.2cm}
\centering
{\fontfamily{qbk}\selectfont
Execute the Rewritten Code
}

\begin{minted}[bgcolor=light-gray, escapeinside=||]{python}
ipython.run_cell(new_source)
\end{minted}
\end{minipage}%
};

\draw [-stealth] (codeSource) -- (PattMatcher);
\draw [-stealth] (PattMatcher) -- (OnlRewr);
\draw [-stealth] (OnlRewr) -- (RunCell);

\end{tikzpicture}
\caption{\system{} overview. \system{} identifies patterns in the source code,
which it rewrites using its rewriter. The optimized version is used only if
certain dynamic checks pass, to ensure correctness.}
\label{fig:system_overview}
\end{figure}


We now present the high-level architecture of \system{}, a rewrite engine that
automatically applies rewrite rules to improve the performance of ad-hoc EDA
workloads.

We designed \system{} with two high-level components. First, \system{}'
\textit{syntactic pattern matcher} matches the input code against the syntactic
patterns the rewrite rules. The second component is a \emph{rewriter}, which
checks the runtime preconditions of the rewrite rules and on success, rewrites
the code to an equivalent, but faster version and executes it. We show a
high-level overview in Figure \ref{fig:system_overview}.

We have several desiderata for \system: it should facilitate applying complex
rewrites automatically with minimal overhead. Further, it should guarantee that
the rewritten code is semantically equivalent to the original code i.e.,
that it is sound, which presents the main technical challenge.

To guarantee soundness, we first formalize the rewrites
(Section~\ref{sec:pandas_rewr_rules}). Second, we accurately describe the
conditions under which a rewrite can be applied, some of which can be quite
involved. For example, some require \system{} to check the form of whole
functions. And finally, most of these conditions can only be checked at runtime,
and checking them at the correct program points requires delicate program
transformations (Section~\ref{sub-sec:rewriter}).

\hfill{}

\begin{figure}
  \begin{subfigure}{\columnwidth}
\begin{minted}[bgcolor=light-gray]{python}
df['A'].sort_values().head(n=5)
\end{minted}
    \caption{Select the 5 smallest elements by sorting first. Extracted from a
    Kaggle notebook \cite{real_nb_sort_values}.}
    \label{fig:sort_values}
  \end{subfigure}
  \hfill
  \begin{subfigure}{\columnwidth}
\begin{minted}[bgcolor=light-gray]{python}
df['A'].nsmallest(n=5)
\end{minted}
    \caption{Select the 5 smallest elements directly. This avoids sorting.}
    \label{fig:nsmallest}
  \end{subfigure}
  \caption{Selecting the 5 smallest elements. By comprehending
  the \code{pandas} API, \system{} is able to recognize that the second version
  is equivalent to, and faster than, the first.}
  \label{fig:example_rewrite}
\end{figure}
\subsection{Pandas Rewrite Rules}
\label{sec:pandas_rewr_rules}




\begin{figure}
    \begin{subfigure}{\columnwidth}
  \begin{minted}[bgcolor=light-gray,escapeinside=||]{python}
@{expr: called_on}.sort_values().head(n=@{Const(int): first_n})
|{\LARGE\color{brown}$\mapsto$}|
@{called_on}.nsmallest(n=@{first_n})
  \end{minted}
      \caption{LHS {\color{brown}$\mapsto$} RHS}
      \label{fig:sort-values-rule-lhs-rhs}
    \end{subfigure}
    \hfill
    \begin{subfigure}{\columnwidth}
  \begin{minted}[bgcolor=light-gray,escapeinside=||]{python}
type(@{called_on}) == pandas.Series
  \end{minted}
      \caption{Preconditions}
      \label{fig:sort-values-rule-preconds}
    \end{subfigure}
    \caption{\revis{An example of a rewrite rule, named \textbf{SortHead}}. If we match the LHS in the source code,
    we can replace it with the RHS only if the preconditions hold (at runtime).}
    \label{fig:sort-values-rule}
  \end{figure}

The abstract form of the rewrite rules \system{} supports can be modeled as transforming a Left Hand Side (LHS) set of statements to Right Hand Side (RHS) set of statements subject to certain preconditions on the LHS.

The most general form of a rule consists of the following:

\begin{enumerate}[leftmargin=25pt]
  \item \textbf{LHS}: A code fragment with some parts that may vary. We call
  those the \textbf{varyingSet}. This LHS must be recognizable with a subtree
  search on the AST.
  \item \textbf{TransformLHS}: An AST transformer that transforms the LHS. We
  call the result the RHS of the rule.
  \item \textbf{RuntimePrecond}: An algorithm for checking the runtime
  preconditions under which using the RHS instead of the LHS is
  semantic-preserving. This is essentially a callback function, with inputs a
  subset of the \textbf{varyingSet}, but evaluated. This callback also gets a
  state snapshot of the interpreter as input. It returns \code{true} or \code{false}.
\end{enumerate}

\paragraph{\textbf{LHS}} We introduce some notation to show the structure of the
LHS. Consider the original code in Figure~\ref{fig:example_rewrite}(a) rewritten
to Figure~\ref{fig:example_rewrite}(b) using the rewrite rule shown in
Figure~\ref{fig:sort-values-rule}. A \code{@\{...\}} entry denotes a varying
part of the rule (i.e., an element of the \textbf{varyingSet}). These parts can
be matched to multiple valid options. Inside the curly brackets, we describe
these valid options using a derivation rule of the Python grammar
\cite{python_ast}. For example, \code{@\{expr\}} denotes that any expression can
appear in its place. For \code{Constant}s, which we refer to as \code{Const} for
short, we optionally specify the type of the constant inside
parentheses.\footnote{We can determine the type of constants from the AST
\cite{python_ast_constants}.} So, \code{@\{Const(int)\}} denotes that any
integer constant can appear in its place. We need to refer to elements of the
\textbf{varyingSet} in other components of the rewrite rule. So, we bind them to
names. For example, the code string \code{df.sort\_values().head()} matches the
LHS of Figure~\ref{fig:sort-values-rule} and \code{called\_on} is bound to (the
string) \code{df}. Everything that is not in \code{@\{\}} should appear as is.
With these in mind, we can read the LHS of Figure~\ref{fig:sort-values-rule} as
matching any Python expression on which \code{sort\_values()} is applied,
followed by \code{head()} with any constant integer as the argument of the
formal parameter \code{n}.

\begin{figure}
  \begin{minted}[bgcolor=light-gray]{python}
"{}.nsmallest(n={})".format(called_on, first_n)
  \end{minted}
      \caption{A sketch of the implementation of the RHS of the rule in Figure~\ref{fig:sort-values-rule}}
      \label{fig:sort-values-rhs}
  \end{figure}
\begin{figure}
  \begin{subfigure}{\columnwidth}
\begin{minted}[bgcolor=light-gray]{python}
def foo(x):
  return True if x['Fare'] > 10 else False
df.apply(foo, axis=1)
\end{minted}
    \caption{A function applied to the whole dataframe.}
    \label{fig:remove-axis-1-orig}
  \end{subfigure}
  \hfill
  \begin{subfigure}{\columnwidth}
\begin{minted}[bgcolor=light-gray]{python}
def foo(x):
  return True if x > 10 else False
df['Fare'].apply(foo)
\end{minted}
    \caption{The function touches only one column so we can increase locality by iterating only that column.}
    \label{fig:remove-axis-1-rewr}
  \end{subfigure}
  \caption{Apply a function to a single column instead of a whole dataframe and increase locality}
  \label{fig:remove-axis-1-ex}
\end{figure}

\paragraph{\textbf{TransformLHS}} This is a function which takes as input
the \textbf{LHS} as an AST and outputs the RHS as an AST. The simplest form is a
function that outputs a constant string, with some elements of
\textbf{varyingSet} (also strings) interpolated. For example, the RHS of
Figure~\ref{fig:sort-values-rule} can be implemented as shown in
Figure~\ref{fig:sort-values-rhs}. The values of \code{called\_on} and
\code{first\_n} are part of the \textbf{varyingSet}. This LHS transformer is
quite simple, but others can be significantly more complex and they can also
depend on \textit{dynamic} information (the transformer we just described does
not as \code{called\_on} and \code{first\_n} are extracted from the text).

\paragraph{\textbf{RuntimePrecond}} The runtime preconditions describe
conditions which have to hold at \emph{runtime} for the original (LHS) and the
RHS to be semantically equivalent. For example, in Figure
\ref{fig:sort-values-rule}, the result of the \code{called\_on} expression that
was matched in the LHS should be a \code{pandas.DataFrame}. The runtime
preconditions implicitly impose an order of evaluation. In this example,
\code{called\_on} must be evaluated first, then the preconditions are checked on
the resulting object, and then this object is used in place of \code{called\_on}
in the RHS. Note that unconditionally evaluating \code{called\_on} is correct
even if the conditions do not hold because it would be evaluated anyway in the
original.

In general, \textbf{RuntimePrecond} imposes some restrictions
which are up to the rule designer and implementer. For example, it should be the
case that the subset of \textbf{varyingSet} used in the preconditions is
evaluated regardless of the result of the precondition check.

In their generic form, the runtime preconditions are also functions that
depend on dynamic information. These can be simple, like the one in
Figure~\ref{fig:sort-values-rule}, or more complicated (i.e., whole
algorithms).

\begin{table*}[t]
  \centering
  \includegraphics[height=0.85\textheight]{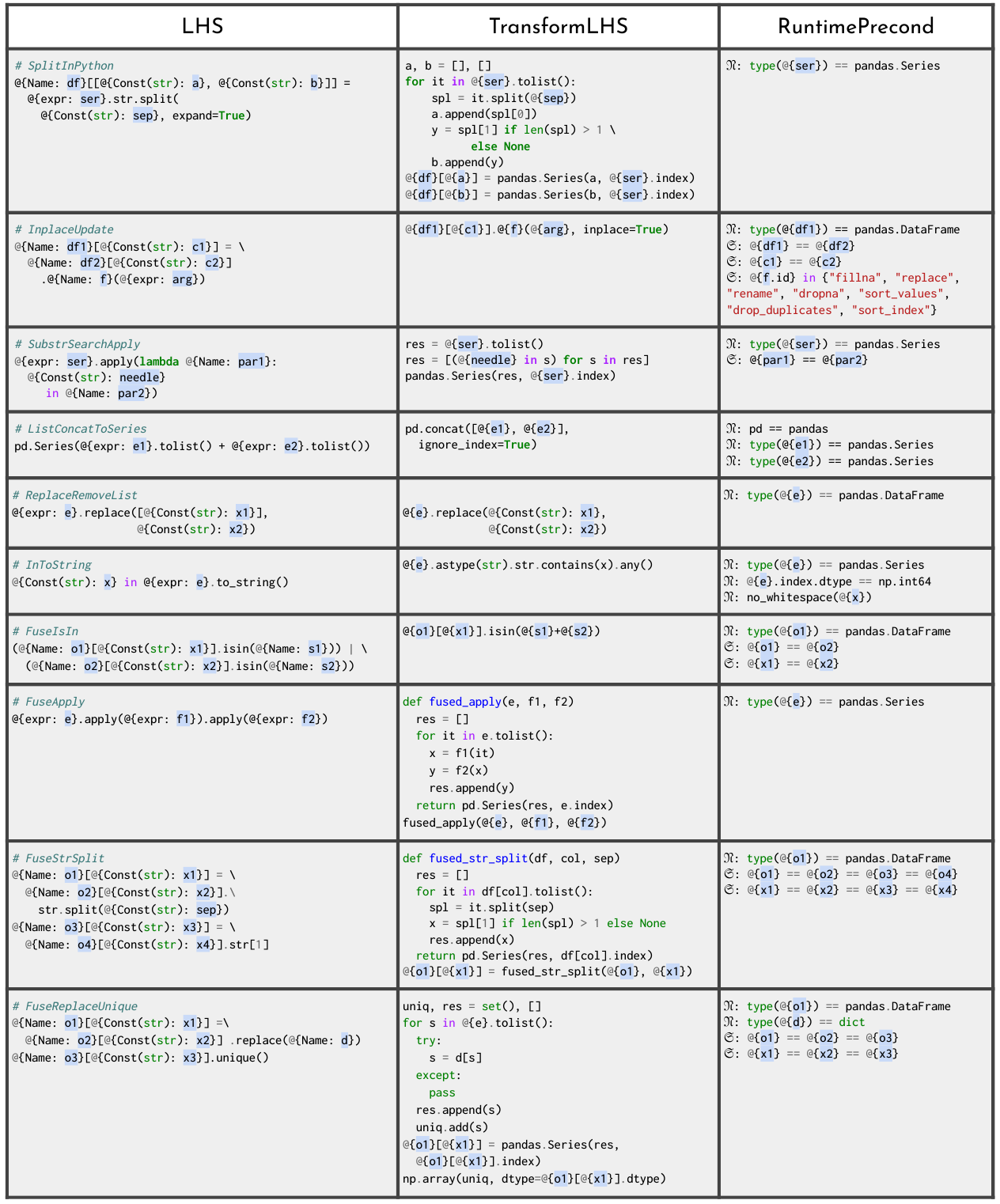}
  \caption{Examples of Rewrite Rules. If any of the LHS's is matched, it can be
  replaced with the corresponding RHS, provided that the preconditions hold. The
  symbol $\mathfrak S$ denotes syntactic preconditions while $\mathfrak R$
  denotes runtime ones. The name of the rule appears as a comment in the LHS
  column.}
  \label{tbl:rewr_rules}
\end{table*}

\revis{Table~\ref{tbl:rewr_rules} shows nine more rewrite rules we use in
\system{}. The first two correspond to the examples in Figure~\ref{fig:split}
and Figure~\ref{fig:concat-with-lists}, respectively. The \code{TransformLHS}'s
are simple transformers which output a paremeterized string, similar
Figure~\ref{fig:sort-values-rhs} which we saw above. Also, we introduce some
notation for syntactic preconditions, which are prefixed by $\mathfrak S$. These
are simple syntactic conditions (i.e., equalities on ASTs) which are part of the
LHS (and which are checked in the pattern matcher). They differ from the runtime
preconditions (\textbf{RuntimePrecond}), which are prefixed by $\mathfrak R$. 

Together with the rules in Figure~\ref{fig:sort-values-rule}, the rule that
achieves the rewrite in Figure~\ref{fig:apply_only_math} (named
\code{ApplyOnlyMath}), \code{RemoveAxis1} (which we discuss below) and the
\code{VectorizedConditionals} rule (which we discuss as a case study in
Section~\ref{sub-sec:ablation}), they make up all the rules used in \system{}.
We note that we leave for future work the formal description of the latter three
rules.}

\begin{figure}
\begin{minted}[bgcolor=light-gray]{python}
@{expr: called_on}.apply(@{Name: func}, axis=@{expr: axis})
\end{minted}
  \caption{RemoveAxis1 rule's LHS}
  \label{fig:remove-axis-1-lhs}
\end{figure}

\paragraph{\textbf{RemoveAxis1}} We will briefly discuss one more rule,
called \code{RemoveAxis1}, whose \textbf{RuntimePrecond} and
\textbf{TransformLHS} components are significantly more complicated. An example
of applying this rule is shown in Figure~\ref{fig:remove-axis-1-ex}. This rule
targets cases where \code{apply()} is applied to a whole \code{DataFrame}. It
checks whether the function passed to \code{apply()} touches only a single
column and if so, it rewrites the code so that the function is applied only to
this column.

We show the LHS in Figure~\ref{fig:remove-axis-1-lhs}. The
\textbf{RuntimePrecond} component, which we do not include here for clarity
purposes, is a whole algorithm which basically checks that only a single column
of the \code{DataFrame} is accessed. The \textbf{TransformLHS} component
replaces all \code{Subscript}'s inside \code{@\{func\}} with just their object
and it removes the \code{axis} argument.

\section{\system{} Rewrite System}

\system{} consists of two main parts: a syntactic pattern matcher and a rewriter
that rewrites the code matched against patterns. We now describe how the two
parts were designed in detail.

\subsection{\system{} Pattern Matcher}
\label{sub-sec:patt-match}

The pattern matcher is responsible for matching a sequence of statements
with the \textbf{LHS} part of any rewrite rule. Whether any \textbf{LHS}
(represented as an AST) matches any part of the source AST, is essentially a
sub-tree search problem. The pattern matcher performs a greedy search and it
returns the first LHS it matches.

To minimize matching overhead, we designed the pattern matcher to be
hierarchical, by factoring patterns based on their commonalities. The common
parts are matched first before hierarchically matching more specific components
of a rule. This eliminates repeatedly matching components that are common to
multiple rules.


Lastly, the pattern matcher needs to be able to match patterns that span
multiple statements. Having a function that matches single-statement patterns by
performing a greedy search, there is another function that matches multiple
statements. The latter function operates on a higher level, viewing
multi-statement patterns as sets of smaller ones. So, for a 2-statement pattern,
if it matches the first part, it then checks the next statement for the second
part.

\subsection{\system{} Rewriter}
\label{sub-sec:rewriter}



When a piece of code is successfully matched with a rewrite rule's
\textbf{LHS}, if there are no runtime preconditions (i.e.,
\textbf{RuntimePrecond} just returns \code{True}), then the rewriter can invoke
\textbf{TransformLHS} on the \textbf{LHS}, and replace the \textbf{LHS} with the
result.

For example, consider the rule in Figure~\ref{fig:sort-values-rule}.
\code{@\{called\_on\}} needs to be evaluated first, let us name the resulting
object \code{res}, then execution needs to \textit{stop}, check the precondition
on \code{res}, and then continue (i.e., evaluating either
\code{res.sort\_values().head()} or \code{res.nsmallest()}, depending on the
precondition result).

This is more difficult than it looks because we do not have arbitrary
control over the evaluation of the code, since we are operating at the source
level. So, we need to \textit{effectively} do the same thing but using
source-level transformations. This is difficult because we are not allowed to
evaluate \code{@\{called\_on\}} twice. The obvious solution is to just evaluate
it once and reuse it.

\begin{figure}
  \begin{minted}[bgcolor=light-gray,escapeinside=||]{python}
test(mod_x(), foo().read_x().sort_values().head())
  \end{minted}
    \caption{A nested expression with global state access.}
    \label{fig:sort-values-nested}
  \end{figure}
\begin{figure}
  \begin{minted}[bgcolor=light-gray,escapeinside=||]{python}
def sort_head(tmp):
  return tmp.nsmallest() if type(tmp) == pd.Series 
    else tmp.sort_values().head()

test(mod_x(), sort_head(foo.read_x()))
  \end{minted}
    \caption{A correct dynamic check with a local binding.}
    \label{fig:sort-values-precond-correct}
  \end{figure}

However, this requires careful orchestration. Consider for example the code in
Figure~\ref{fig:sort-values-nested}. The evaluation-and-reuse should happen
\textit{exactly} where the original sub-expression (involving
\code{sort\_values()}) appears. Otherwise (e.g., if we save it in a variable by
adding a statement above), it is possible that \code{read\_x()} will read a
stale value.

To solve that in general, we need a local binding of the evaluation of
\code{@\{called\_on\}} \footnote{In the style of a \code{let} expression in
OCaml.}. However, Python does not have local bindings, so the workaround is to
create a function and call it at the place of the original expression, as in
Figure~\ref{fig:sort-values-precond-correct}. The local binding here is the
binding to the function's parameter.

\paragraph{\textbf{Dynamic RHS}}

Observe that in the solution we just mentioned, to create the function
\code{sort\_head()}, we need to know the RHS a-priori. This is true when
\textbf{TransformLHS} does not depend on dynamic information and thus we can
``run'' it offline. This is the case for the rule in
Figure~\ref{fig:sort-values-rule}. However, this is not the case for the
\code{RemoveAxis1} rule, because it depends on the body of \code{@\{func\}},
which is not known offline. To implement such rules, all of which involve
\code{apply()} and which face the same problem, we just overwrite
\code{apply()}. In the overwritten body, we invoke \textbf{RuntimePrecond},
which depends on the body of \code{@\{func\}}, which is however available
because it is passed as an argument. If the preconditions pass, we rewrite the
body on the fly and invoke it appropriately. For example, in the case of
\code{RemoveAxis1}, we rewrite the body as described earlier and we call
\code{self[theOneSeries].apply(new\_body)}. Note that \code{self} is the
evaluated \code{@\{expr\}}. We never see this \code{@\{expr\}}, but we know this
fact because \code{self} is bound to the object on which (the overwritten) \code{apply()}
gets called.



\section{Implementation}
\label{sec:Implementation}

\subsection{IPython Integration}

To work automatically, \system{} currently depends on IPython
\cite{ipython_org}, which is an enhanced Python interpreter. This implies that a
current limitation of our implementation is that \system{} is not automatic in
standard Python. In practice, this is not a problem because the dominant
platform for the notebooks we target is the IPython notebook (usually accessed
through Jupyter \cite{jupyter}), which requires IPython. However, \system{} can
still be used as a library even with a standard Python interpreter. In the rest
of this section, we will assume that \system{} is running on top of IPython.

An IPython notebook consists of a collection of code snippets called
\emph{cells}. Each cell can be executed individually, which is commonly done in
interactive EDA workloads. The goal of our implementation to invoke \system{}
automatically before a cell is executed, and rewrite it automatically, on the
fly. A key feature of IPython that allows us to do that is the input transformer
\cite{ipython_input_transformer}. This allows us to register a function that
runs before a cell gets executed. That function gets the cell content as a
string and returns a new string which becomes the new cell. Our input
transformer just inserts a call to \system{}' main routine, with the original
cell passed as an argument. \system{} then potentially rewrites and finally
executes the cell. This is all automatic; the user just needs to import
\system{}.

\system{}' main routine gets the cell code as a string, which it first parses as
an AST. For that, we use the Python \code{ast} library \cite{python_ast}, which
parses Python code. This implies a limitation because cells can contain invalid
Python syntax (but valid for IPython, e.g., magic functions), which this library
will not handle. This did not cause serious problems in practice but we hope to
fix it in the future.

An important detail is that \system{} runs on the \emph{same} IPython instance
as the notebook, having access to the same namespace as the underlying cells.
This is necessary because \system{} needs to inspect dynamic information
like names, types, function objects, etc.

\subsection{Crossing Library Boundaries}

It might seem that we could achieve the same optimizations simply by modifying
the \code{pandas} library. In fact, for some rules, the implementation is
effectively a replacement of \code{pandas} routines with our own. For example,
\code{RemoveAxis1} is implemented by overriding \code{pandas}'s \code{apply()}.
This way, we get both the \code{@\{func\}} and the \code{@\{called\_on\}}
components easily, without needing to do source-level transformations (like the
ones we described in Section~\ref{sub-sec:rewriter}). The former is accessed
through the \code{self} implicit argument, and the latter through the
\code{func} argument to \code{apply()}.

However, this approach does not suit all cases. First, in some cases, it is
simply easier to operate as an external rewriter. For example, the rule in
Figure~\ref{fig:sort-values-rule} can be applied if \code{sort\_values()} is
followed by \code{head()}. If we overwrite \code{sort\_values()}, we cannot know
what happens with its result. If we overwrite \code{head()}, we cannot know how
the object it is applied to came to be. There are ways to work around these
limitations and one popular one is lazy execution, which has been employed for
similar purposes \cite{best_effort_lazy} (Modin also employs it).

In summary, in lazy execution, we do not execute the code but rather we log which
functions have been called. After a call chain is evaluated, the result is a
computation graph that captures the whole computation. We can evaluate it in
the trivial way (e.g., actually calling the functions in sequence), or in an
optimized way (e.g., by calling \code{nsmallest}).



The problem, however, is that this requires a lot of bookkeeping to know when
exactly to evaluate the computation graph, to hide from the user the fact that
the functions do not return the type the user expects them to return, to build
this computation graph, etc. In other words, lazy execution is a hack around the
fact that we do not see all the computation and we are just trying to
reconstruct it from inside a library. With an external rewriter, this is
trivially solved because we just view all the code. So, we just apply well-known
code transformation techniques.

But the most important benefit of an external rewriter is that it can
effortlessly rewrite across representations. For example, it would be nearly
impossible to support the rewrite shown in Figure~\ref{fig:concat-with-lists}
with lazy execution by keeping the API intact because \code{tolist()} is
supposed to actually return a Python list, thereby moving us away from the
library's control. So, the lists will unavoidably get concatenated in pure
Python. But an external rewriter just views the code and it can trivially
perform the rewrite.




\section{Evaluation}
\label{sec:Evaluation}

\subsection{Experimental Setup}
\label{sub-sec:exp-setup}

All the experiments, except if mentioned otherwise, were performed on a system
with a 12-core AMD Ryzen 5900X, 32GB of main memory, Samsung 980 PRO NVMe SSD and Ubuntu
22.04.1 LTS.

\paragraph{\textbf{Benchmark}}

Our goal was to evaluate \system{} on real workloads and so we picked notebooks
from Kaggle. We chose Kaggle as it is a popular repository for data science
workloads and it also contains both the data and notebooks used. The overarching
hypothesis that we want to validate is that a rewrite system like \system{} can
offer substantial speedups on real-world notebooks, through rewriting, with
minimal slowdowns, minimal memory consumption and disk usage, and without
changing the API.

In this work, we focus on ad-hoc EDA, \code{pandas}-heavy workloads. In order to
find such notebooks, we chose notebooks randomly from Kaggle subject to the
following conditions:
\begin{itemize}
    \item At least 50\% of static function calls are \code{pandas} calls
    \item Using datasets of size approximately 2GB or less
\end{itemize}

We chose the first criterion because we focus on EDA notebooks. In particular,
many of the notebooks we excluded focused on machine learning and plotting,
which are out of scope for this work. In the notebooks we picked, we disabled
such code for our evaluation.

Our second criterion was to filter out notebooks that were already
hand-optimized. These notebooks typically operated on large datasets.
Optimization is necessary in this setting as Kaggle has resource constraints
(both computational and memory). However, hand-optimization requires significant
effort. \system{} is an automatic and transparent system and we want to evaluate
its effectiveness without users having to expend that effort.

For the datasets that were significantly lower than 2GB, we replicated them so
that they reach at least several hundred MBs (otherwise our measurements would
be dominated by noise). Also, we modified any notebook that used a sample/subset
of the dataset to instead operate on the full dataset.



We sampled 20 notebooks satisfying our criteria. There are rewrite opportunities
in 10 of these 20 notebooks, which we coded in \system{}. We focus on these 10
notebooks in our evaluation. We further executed \system{} on the remainder of
the notebooks where no patterns were matched to study \system{}' overhead. We
describe further experiments that include all 20 notebooks in an extended
version of this manuscript \footnote{Not cited to preserve anonymity}. We
compare \system{} with \code{pandas} (version 1.5.1) and \code{modin}
\cite{modin} (version 0.17.0).

\subsection{End-to-End vs Pandas}
\label{sub-sec:vs-pandas}

\begin{figure*}[t]
    \centering
    \includegraphics[width=\textwidth]{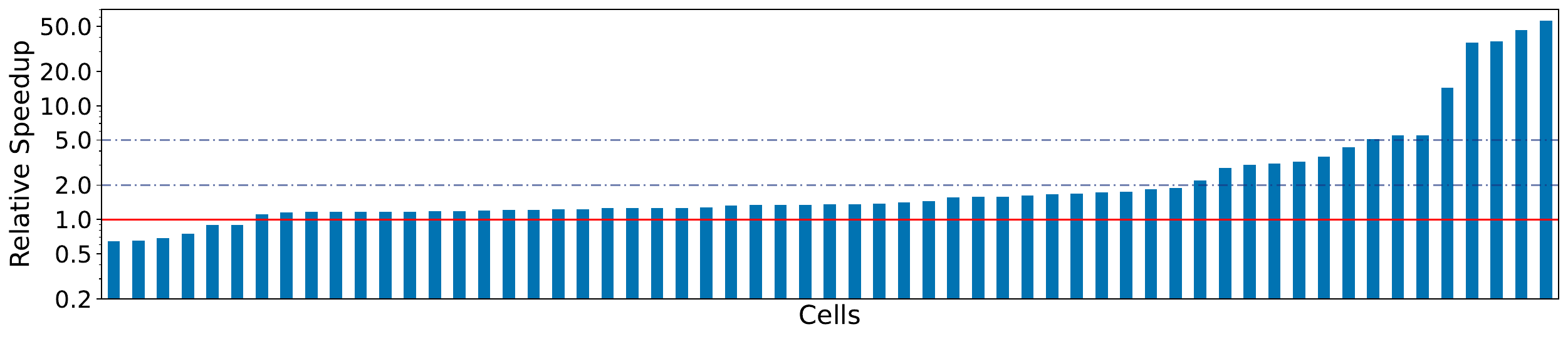}
    \caption{Cell-level relative speedups (excluding cells that originally ran
    for less than 50ms and also all the cells that got a speedup or slowdown
    within the \revis{0.1$\times$ range}). Again, \system{} provides significant speedups by up
    to 57$\times$. There are also slowdowns, which are not substantial (see
    Figure~\ref{fig:cells-slowdowns}).}
    \label{fig:cell_level}
\end{figure*}
\begin{figure}[ht]
    \centering
    \includegraphics[width=0.7\columnwidth]{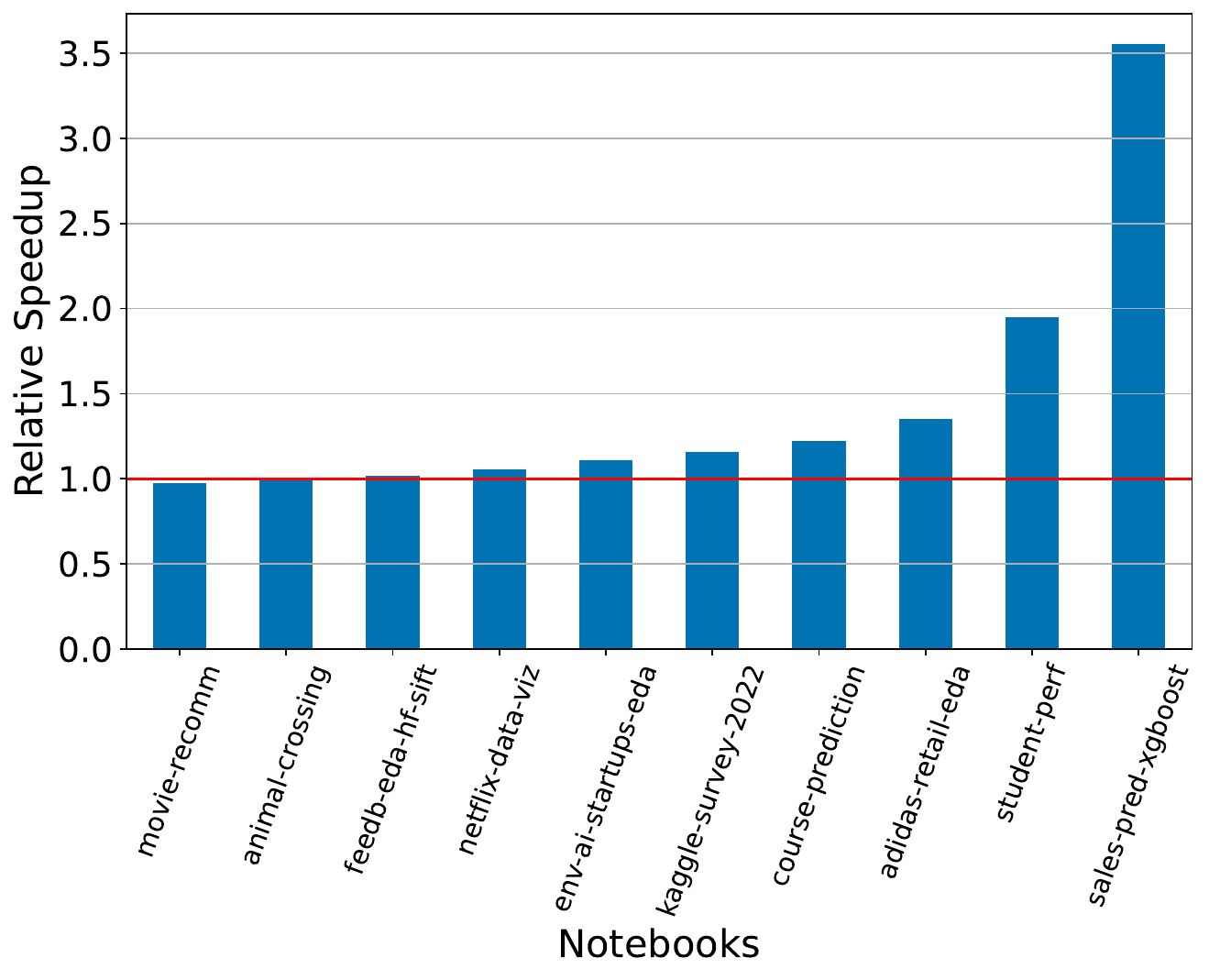}
    \caption{Relative speedups on whole notebooks. \system{} speeds up notebooks
    by up to 3.6$\times$ while not slowing down any notebook by more than \revis{1.03$\times$}. }
    \label{fig:nb_level}
\end{figure}

We first investigated whether \system{} can accelerate cells and notebooks
compared to standard \texttt{pandas}. To do so, we ran each sampled notebook
with and without \system{}. We ran 10 trials each and measured execution time at
the cell level. Our primary metric was the speedup of cells and notebooks with
\system{} compared to standard \texttt{pandas}. We report the geometric mean of
the speedups.

\paragraph{\textbf{Per-Notebook Speedups}} We show per-notebook relative
speedups in Figure \ref{fig:nb_level}. As shown, \system{} can provide
substantial speedups \emph{at the notebook level} of up to 3.6$\times$. Overall,
\system{} provides significant speedups in half of the notebooks (five) and
moderate speedups in one other notebook. We emphasize that these notebooks were
selected randomly from Kaggle, showing the applicability of \system{}.

Furthermore, \system{} does not significantly slow down any notebook, with a
maximum slowdown \revis{1.03$\times$}. \system{} rewrites cells
in these notebooks but it does not achieve speedups.

\paragraph{\textbf{Per-Cell Speedups}} We show per-cell speedups in Figure
\ref{fig:cell_level}. For clarity, we excluded cells that run for fewer than
50ms in the original version and excluded all speedups and slowdowns when run
with \system{} within \revis{0.1x} the original cell runtime.

As shown, \system{} can achieve per-cell speedups of up to 57$\times$. The cell
with the highest speedup is matched by the pattern shown in
Figure~\ref{fig:apply_only_math}. The second largest speedup is due to the
pattern \revis{\code{Vectorized} \code{Conditionals}, which is} discussed in
Section~\ref{sub-sec:ablation}. The majority of cells we consider are improved
by \system{}. The maximum slowdown in an individual cell is
\revis{1.56$\times$}. In general, the cells that have the highest slowdown are
fast cells, i.e., those already within interactive latencies, both before and
after rewriting.

\paragraph{\textbf{Overhead of \system{}}} We further investigated the cause of
slowdowns. We first measured the overhead of deploying \system{} (on all 20
notebooks). We find that \system{} never has an overhead of more than \emph{22
ms} with a geometric mean overhead of 0.41ms.

\begin{figure}[ht]
  \centering
  \includegraphics[width=0.7\columnwidth]{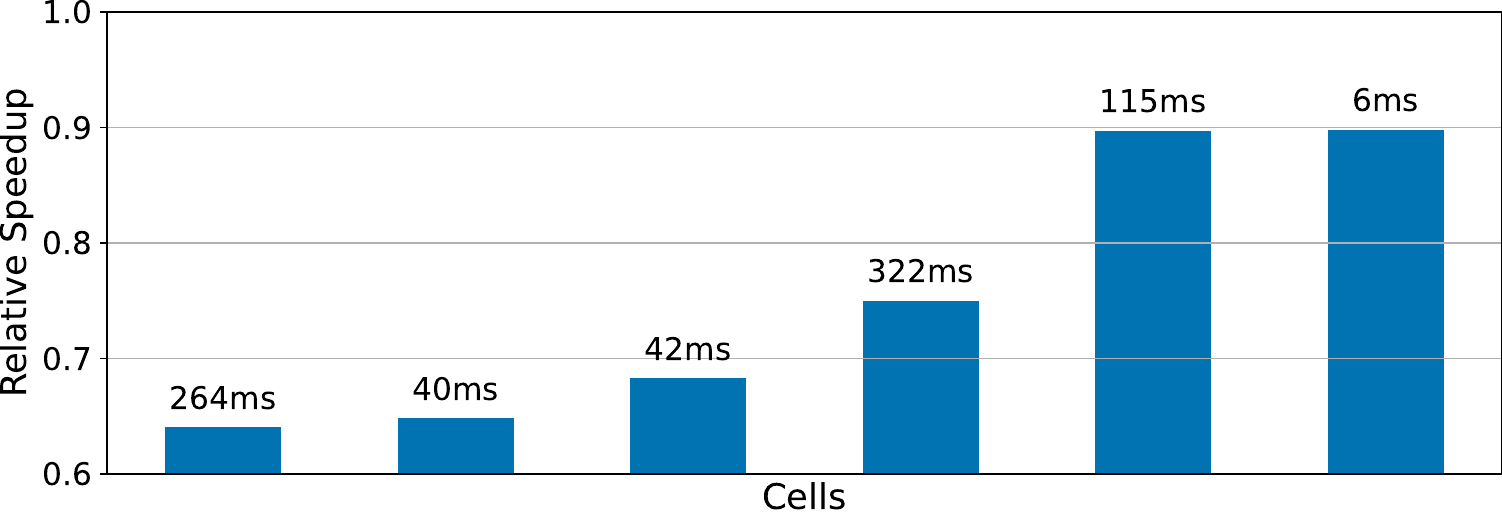}
  \caption{The subset of cells from Figure~\ref{fig:cell_level} that got slowed
  down. Above the bars, we show the \textit{absolute} slowdown. The slowdowns are
  within interactive latencies (i.e., less than 300ms), with the maximum
  overhead of \system{} being 23ms.}
  \label{fig:cells-slowdowns}
\end{figure}

However, in addition to the overhead from deploying \system{}, \system{} may
also cause downstream effects. We find that in some cases, cells that are not
modified by \system{} can experience degradations in performance. The highest
magnitude of those appear only in notebooks where \system{} rewrites cells.
Because of this, and because some of these slowdowns are much larger than any
overhead that \system{} can cause, we hypothesize that rewriting is not the
cause of the slowdown. Rather, it seems that the rewritten version of a cell,
while faster than the original version of this same cell, causes a slowdown in
another cell of the same notebook. Nonetheless, these slowdowns are not
substantial. In Figure~\ref{fig:cells-slowdowns} we show only the cells from
Figure~\ref{fig:cell_level} that get slowdowns along with the absolute slowdown.
That figure shows that even when the relative slowdown is large, the absolute
slowdown is below interactive latency times (i.e., below 300ms).

\begin{figure*}[ht]
  \centering
  \includegraphics[width=\textwidth]{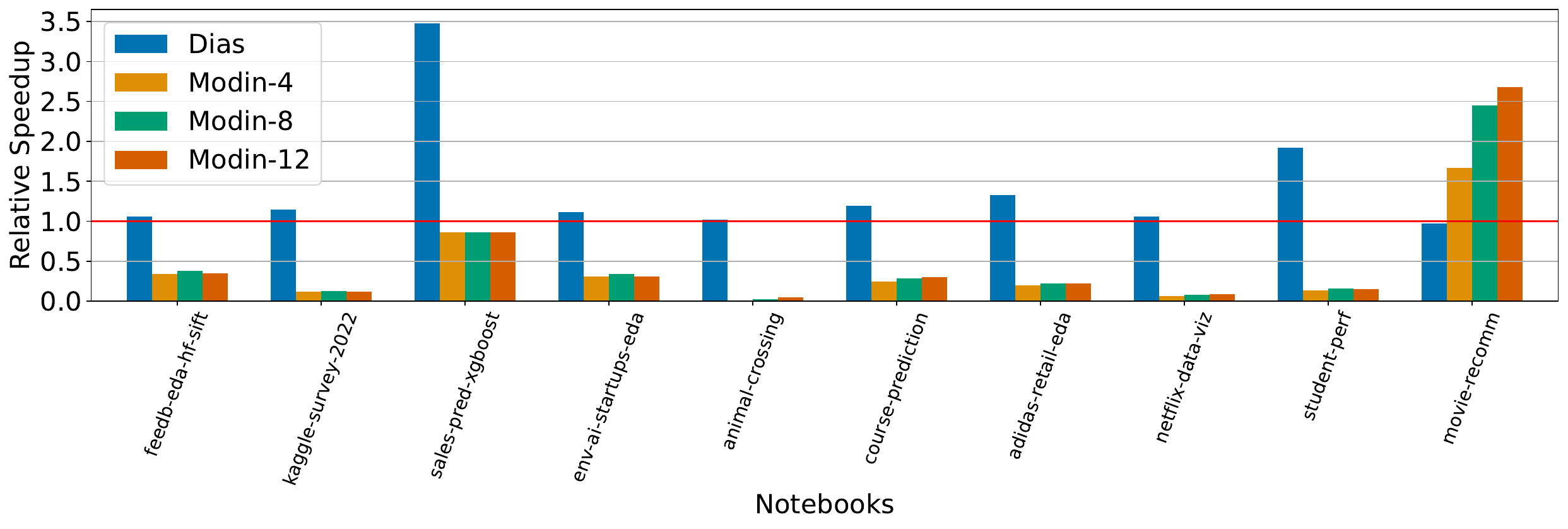}
  \caption{Comparing \system{} with \code{modin} \cite{modin}. \system{} is
  faster for 9 out of 10 notebooks (Up to 27.1$\times$ faster with 4.9$\times$
  geometric mean). \code{modin} is, in many cases significantly, slower than the
  original for these 9 notebooks. For the one notebook where \system{} is
  slower, it is no more slower than \revis{1.03$\times$} compared to the original.}
  \label{fig:modin_nb_level}
\end{figure*}

\subsection{Comparison with Modin}

We compare \system{} with \code{modin} \cite{modin} (using Ray as the underlying engine
which is the default). We chose \code{modin} because it enjoys wide adoption and is
supposed to be a drop-in replacement for \code{pandas}.

We focus on deploying \code{modin} on a single server as this is the setting we focus
on in this work. Unfortunately, we find that deploying \code{modin} in this setting is
difficult for two reasons: excess memory utilization and lack of support for the
full \texttt{pandas} API.

For the notebooks we consider, \code{modin} consumes substantially more memory
resources than standard \texttt{pandas}. Even when using a powerful AWS server,
the AWS \texttt{c5.24xlarge} with 96 vCPUs and 192 GB of RAM, \code{modin} was
unable to execute five of the ten notebooks we consider. As a result, we modify
the default \code{modin} settings to execute on 4 to 12 cores depending on the
notebook and we also had to reduce the dataset replication on 3 of the 10
notebooks. With these modifications, we are able to run the notebooks with
\code{modin}, using our original setup.

We further find that \code{modin} does not support 100\% of the \texttt{pandas} API. As
a result, we could not run two of the ten notebooks. We changed the impeding
snippets to ones that are functionally close. Given our new setup, we compared
\code{modin}, \system{}, and vanilla \texttt{pandas}.

\begin{figure}[ht]
  \centering
  \includegraphics[width=0.7\columnwidth]{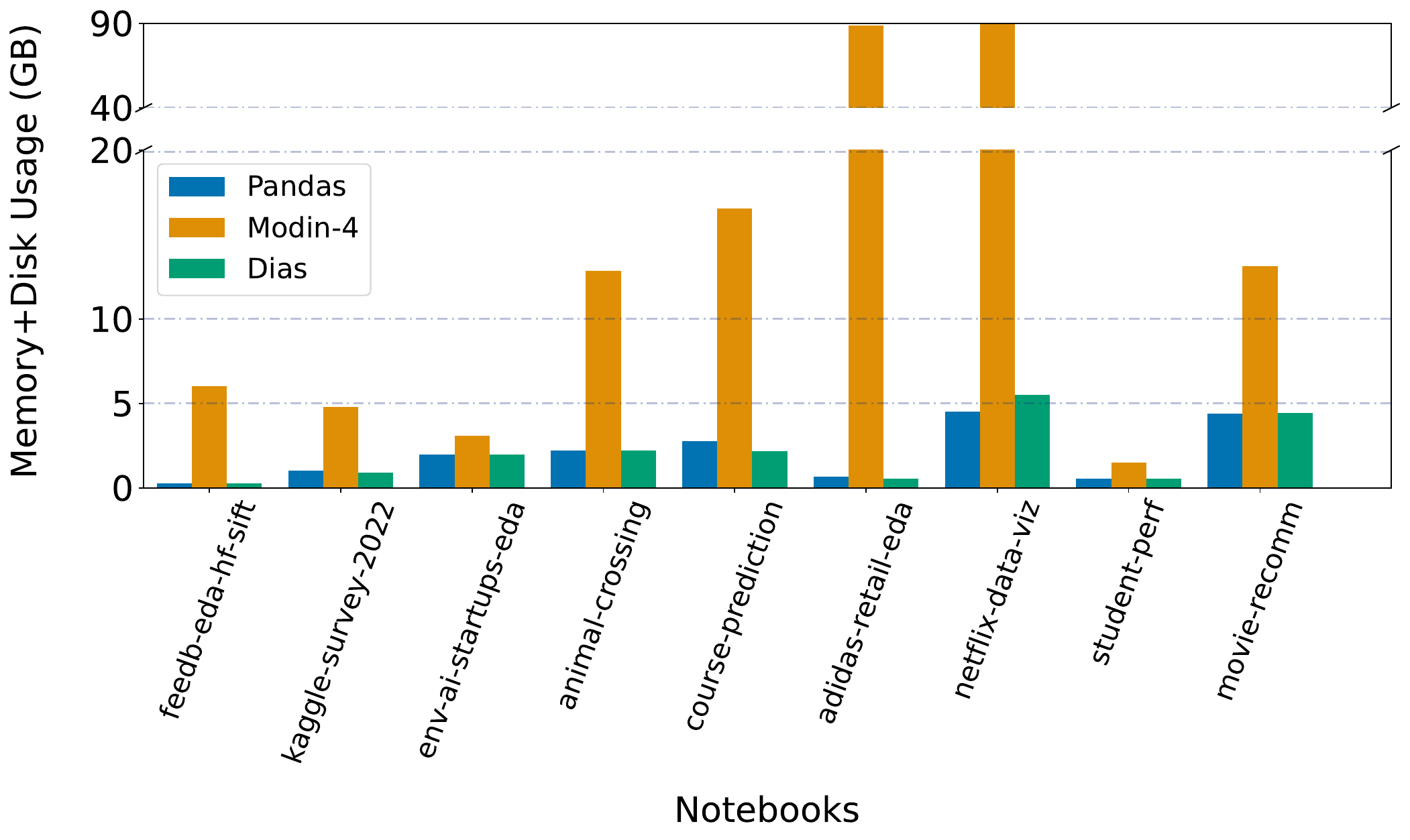}
  \caption{RAM and disk usage comparison in \code{modin}, \system{} and \code{pandas}.
  \system{} and \code{pandas} do not use the disk and they use almost the same
  amount of RAM in all cases. \code{modin} uses the RAM and disk aggressively,
  surpassing the 80GB threshold for a notebook where \code{pandas}/\system{} use
  less than 5GB.}
  \label{fig:modin_nb_mem}
\end{figure}

As shown in Figure \ref{fig:modin_nb_level} \footnote{\system{}' results in
Figure~\ref{fig:modin_nb_level} look slightly different from those in
Figure~\ref{fig:nb_level}, even though the same notebooks are used. This is
because of the changes we had to perform on some of the notebooks (i.e., less
replication and API changes) to run them with \code{modin}.}, \code{modin}
\emph{slows down} 9 of the 10 total notebooks we consider compared to vanilla
\texttt{pandas}. It speeds up one notebook which is dominated by a call to
\code{apply()}, which \code{modin} is able to parallelize. As witnessed in this
notebook, one advantage of \code{modin} is that it can scale with the
availability of more hardware resources in cases where it can parallelize.
\system{} does not enjoy such scaling benefits. However, we find that
\code{modin} cannot parallelize the majority of the notebooks we consider
diminishing any scaling benefits.
Overall, \system{} is up to 27.1$\times$ faster than \code{modin} (4.9$\times$
geometric mean) for whole notebooks.

\revis{Note that if we include all 20 notebooks in our evaluation, \code{modin}
slows down \textit{all} 10 new notebooks (19/20 $\rightarrow$ 95\%). Also, for
individual cells, \system{} reaches speedups up to 1909$\times$ compared to
\code{modin}.}

We further show that \code{modin} uses memory resources (RAM and disk)
aggressively, with results in Figure \ref{fig:modin_nb_mem} \footnote{The only
way we found to measure \code{modin}'s memory consumption somewhat reliably was
using \code{ray memory}, which however was still unreliable and very slow to
query. We could not obtain memory measurements for 1 notebook.}. When deploying
\code{modin} exclusively across multiple servers, it is generally acceptable to
use all the available hardware resources. However, many of the users of ad-hoc
EDA workloads have limited hardware resources, further highlighting the
deployment issues with \code{modin}. \revis{In fact, if we run \code{modin}
across all 20 notebooks, it consumes up to 250GB when \code{pandas} and
\system{} consume less than 5GB}. Note that \system{}, (like \code{pandas}), makes no
use of the disk.


\revis{
\subsection{An Estimate of \system{}' Hit Rate}

In this section we take a closer look at the hit rate of \system{}. In particular, we
wish to answer the question: How frequently does \system{} rewrite a cell?

For our analysis, we utilized a subset of KGTorrent \cite{kgtorrent}, a dataset
of Jupyter notebooks harvested from Kaggle. This dataset contains notebooks that
are outside the scope of our work, like notebooks which do not use \code{pandas}
at all. After filtering such notebooks out, we extracted \textit{a total of
8,853 notebooks and 177,272 cells}. We cannot run these notebooks because: (a)
we found no reliable way to automatically download a notebook's datasets and (b)
even with a dataset, many notebooks still don't run without manual modification.
Therefore, we perform a static analysis. In particular, we run the pattern
matcher over the notebooks and when it matches a rule, we consider it a "hit".
\system{}' hit rate is 3.2\% (5,586 cells) across cells and 27.1\% (2,395
notebooks) across notebooks.

In our evaluation, we used 20 notebooks with 652 cells. \system{} rewrites in 2\% of
all cells (5\% if we discard cells that run for less than 50ms) and 50\% of
notebooks. Because 652 cells is a large sample, and because it agrees with the
static analysis, we contend that \system{}'s hit rate is close to 2-3\%.

This result is significant considering that \system{} currently uses only 12
rules, which is a small set of rules for an automatic rewriter. Production
rewriters have hundreds of rules. For example, TensorFlow r1.14 includes 155
rewrite rules \cite{taso} (which are also simpler), developed over a long period
of time by many engineers and totalling around 53 thousand lines of code. We
emphasize that the novelty of \system{} is in the system, not the specific rules
we happen to use at this snapshot. With \system{}, we wish to encourage such a
development of rules for ad-hoc EDA workloads.

}

\subsection{Understanding \system{}' Performance}
\label{sub-sec:ablation}

To understand the performance gains of \system{}, we discuss two case studies in
detail.



%

\begin{figure}
  \begin{subfigure}{\columnwidth}
\begin{minted}[bgcolor=light-gray]{python}
def foo(row):
  if row['A'] == row['B'] and row['A'] < row['C']:
    return 'X'
  elif row['A'].startswith('Y'):
    return 'Y'
  elif row['B'] in ls:
    return 'Z'
  else:
    return 'NA'

df.apply(foo, axis=1)
\end{minted}
    \caption{Original \code{pandas} \code{apply()}. It processes each row
    sequentially, using the interpreter.}
    \label{fig:apply_vectorized_orig}
  \end{subfigure}
  \hfill
  \begin{subfigure}{\columnwidth}
\begin{minted}[bgcolor=light-gray]{python}
conditions = [
  (df['A'] == df['B']) & (df['A'] < df['C']),
  df['A'].str.startswith('Y'),
  df['B'].isin(ls)
]
choices = [
  'X', 'Y', 'Z'
]
np.select(conditions, choices, default='NA')
\end{minted}
    \caption{Vectorized execution using \code{numpy.select()}}
    \label{fig:apply_vectorized_rewr}
  \end{subfigure}
  \caption{VectorizedConditionals: Vectorized \code{apply()} with conditions, which can be hundreds of
  times faster \cite{pygotham_apply_vectorized}. However, performing this
  rewrite automatically is challenging.}
  \label{fig:apply_vectorized}
\end{figure}

\paragraph{\textbf{Vectorized Conditionals}}
We further study two case studies, starting with \revis{a rule named \code{VectorizedConditionals}}.

We show an example of rewriting a \texttt{pandas} \code{apply()} function with
\code{numpy}'s \code{np.select()} in Figure~\ref{fig:apply_vectorized}
\cite{pygotham_apply_vectorized}. Both versions output a certain value per row
based on some conditions. The second one gives many-fold speedups, 36$\times$ in
our evaluation and up to 380$\times$ in other situations
\cite{pygotham_apply_vectorized}, mainly due to the use of vectorized execution.

To do this rewrite, \system{} checks that the function \code{foo} contains only
an \code{if-else} chain and the conditions are such that we can translate them
to equivalent that apply to whole columns (for example, we cannot translate
\code{if bar()} in some random function \code{bar}). Also, the return
values should be such that can be converted to numpy arrays. The constant
\code{'X'} is such a value but if it were \code{bar(row['A'])}, we would not, in
general, be able to translate it.

Verifying these conditions is not the only tricky part; producing the rewritten
version can be challenging too. For example, the original uses Python's
\emph{logical-AND} (i.e., \code{and}) to compare elements, but we need to use
Python's \emph{bitwise-AND} (i.e., \code{\&}) when translating to \code{pandas} and the
parentheses around the two sides are required. Similarly, a condition like
\code{a in ls} needs to be translated to a call to the \code{pandas} \code{isin()}
function. These are subtleties of rewriting that can be easily missed if we
carry it out manually. Besides leading to bugs, they require extensive
knowledge of \code{pandas}.

As explained in Section~\ref{sub-sec:rewriter}, these checks, and the rewriting,
cannot be performed a priori because the code of \code{foo} might not be
available yet at the start of the cell. Thus, the rewriter employs on-demand
dynamic checking.

Finally, if the user changes \code{foo} such that it does not abide to the
above conditions, the rewriter cannot perform the rewrite. At the same time,
however, the original code remains intact. Thus, the code will never be slower
than the original. Moreover, had the user performed the rewrite by hand, they
would have to convert it back to the \code{apply()} version, an effort
disappears with the rewriter.

\begin{figure}
  \begin{minted}[bgcolor=light-gray]{python}
arr = df['C'].values
n = len(arr)
res = np.empty(n, dtype=arr.dtype)
for i in range(n):
  spl = arr[i].split(',', maxsplit=1)
  res[i] = spl

df_temp = pd.DataFrame(res, columns=['a', 'b'])
a = res['a']
b = res['b']
  \end{minted}
      \caption{\code{Series.str.split()} implementation (Simplified)}
      \label{fig:pandas_split_simplified}
  \end{figure}

\paragraph{\textbf{Translating to Pure Python}} We present a case study of a
non-intuitive result: translating an ``optimized'' \code{pandas} call to pure
Python, \revis{as shown in Figure~\ref{fig:split} (In \system{}, this is covered by the rule
\code{SplitInPython}; see Table~\ref{tbl:rewr_rules})}. In general, users expect
\code{pandas} to be more efficient than pure Python since \code{pandas} uses
vectorized, native code, while also avoiding the interpreter, when possible.

However, \code{.str.split()} is a \emph{string} operation and these cannot in
general be vectorized by \code{numpy}. So, a call to \code{.str.split()}
reaches a standard Python loop to carry out the operation
\cite{pandas_map_infer_mask}.

We would then expect the \code{pandas} version to be in par with our version. We have
to look more closely to understand the discrepancy. In
Figure~\ref{fig:pandas_split_simplified}, we show a simplified version of
\code{.str.split()}'s implementation. Specifically, the important thing is that
in the loop, we gather a collection of (2-element) \emph{lists} in \code{res}
(\code{res} is a \code{numpy} array but it could be any container without much
difference in performance; e.g., it could be a list. The important thing is what
it stores.). Then, we create our two results, our two \code{Series} (via
creating a \code{DataFrame}, but the particular way of doing it is irrelevant). In
particular, we split these lists ``vertically'' and in half so that all the first
elements of the lists create the Series \code{a} and all the second elements
create the Series \code{b}.

One should contrast this with our rewritten version. There, we create only two
lists (\code{a} and \code{b}). At every iteration of the loop, we create one
list, the result of \code{split()}, append the individual elements to \code{a}
and \code{b} and then \emph{throw it away}. Notice that in the \code{pandas}
version, the result of \code{split} has to be saved. So, while on the
surface, the two loops allocate the same number of lists, in our version, the
same space can be reused for every iteration.

Finally, we convert \code{a} and \code{b} (both lists of strings) to
\code{Series}. Under the hood, a list of strings is a contiguous block of memory
in which every element is a pointer to the string. A \code{Series} of strings is
also a contiguous block of memory in which every element is a pointer to a
string. So, the conversion from the one to the other is cheap. However, in the
\code{pandas} version, the elements are stored together in lists ``horizontally'',
but we want to store them together ``vertically'' (if we imagine a matrix where
every row is a list coming from \code{split}), which is expensive.

In this example, the rewriter enables us to optimize a library \emph{without
changing the library}. As we have explained earlier, the rewriter can cross
library boundaries and thus it can optimize across Python, \code{pandas} and \code{numpy},
without the need to provide custom versions of these libraries.

\revis{
\subsection{Comparing Various Dataframe Libraries}
\label{sub-sec:cmp-libs}

To further understand how \code{modin} and other dataframe libraries perform on ad-hoc
EDA workloads, we perform a series of targeted experiments using common patterns
we have found in such workloads. In addition to studying \code{modin}, we also
study three other common dataframe libraries: \code{dask} \cite{dask} (version
2022.12.1), Koalas \cite{koalas_web} (version 0.32.0), and PolaRS
\cite{polars_web} (version 0.7). \code{dask} is another widely adopted parallel
dataframe library with a slightly different API from that of \code{pandas}. Koalas
implements the \code{pandas} API over PySpark \cite{pyspark_web}.  PolaRS
\cite{polars_web} is a \code{pandas} replacement (using Rust under the hood), which,
however, has a different API.

\minihead{Setup and Dataset} We use a \texttt{c5.24xlarge} AWS instance with 96 vCPUs and
192 GiB of RAM. We use 12 vCPUs for \code{modin}, \code{dask}, Koalas and
PolaRS. The dataset used is the NYC Yellow Taxi Dataset 2015 - January
\cite{nyc_taxi_dataset} (except for one case mentioned below) with a size of
around 1.8GB. We picked this dataset because (a) it is large (the subset we use
is the largest we could run the experiments with, using the libraries mentioned,
on this machine) and these libraries specialize in large datasets and (b) it has
been used in previous work \cite{modin} and in multiple notebooks throughout the
Internet \cite{nyc_taxi_kaggle}.

\begin{figure}
  \begin{subfigure}[t]{\columnwidth}
\begin{minted}[bgcolor=light-gray]{python}
df['pickup_longitude'] + df['pickup_latitude']
\end{minted}
    \caption{Example of a column-wise operation: Add Two Series Element-Wise}
    \label{fig:np_sum}
  \end{subfigure}
  \hfill
  \begin{subfigure}[t]{\columnwidth}
\begin{minted}[bgcolor=light-gray]{python}
np.where(pandas_df['A'] < pandas_df['B'], 10, 20)
\end{minted}
    \caption{Example of interacting with \code{numpy}: Vectorized conditional}
    \label{fig:np_where}
  \end{subfigure}
  \caption{Examples of operations performed with the \code{pandas} alternatives.
  These are pretty fast with \code{pandas}.}
  \label{fig:alt_ops}
\end{figure}

\minihead{Operations} We tested several common patterns found in \code{pandas}
workloads, and which are expected to be pretty fast with \code{pandas}
\footnote{For example, \code{pandas} columns are stored as \code{numpy} arrays
and many \code{pandas} operations use \code{numpy}. So, we expect interacting
with \code{numpy} to be quite efficient.}. In particular, we tested column-wise
operations, interacting with \code{numpy}, and an iterative access of individual
elements. Figure~\ref{fig:alt_ops} gives examples of the former two.
Figure~\ref{fig:for_loop} gives an example of the latter.

\begin{table}[t]
  \centering
  \includegraphics[width=\columnwidth]{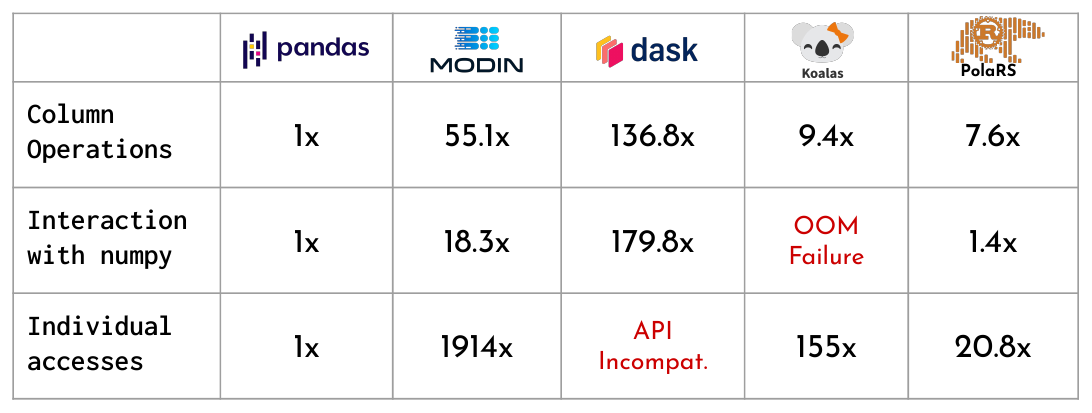}
  \caption{\textit{Slowdown} when running common \code{pandas} operations with
  \code{pandas} alternatives. OOM Failure is out-of-memory failure and API
  incompatibility means that an API we used is not supported. As the results
  show, the current alternatives are not suited to ad-hoc, EDA workloads, both
  in terms of efficiency and usability.}
  \label{tbl:pandas_alt}
\end{table}

\paragraph{\textbf{Results and Discussion}} A subset of our results (for a
single example of each category) is shown in Table~\ref{tbl:pandas_alt}. As is
evident, bulk-parallel dataframe libraries like \code{modin}, \code{dask} and
Koalas, are not well suited for ad-hoc EDA workloads. We should note, however,
that we observed that PolaRS was significantly more judicious with the resources
compared to the other three (especially for memory), and its slowdowns are much
smaller. However, it can still cause considerable slowdowns (e.g., with the
iterative element access) and has a considerably different API. For example, the
\code{pandas} snippet \code{df['A'] = 1} is translated to 
}

\begin{minted}[bgcolor=light-gray]{python}
df = df.with_column(pl.lit(1).alias('A'))
\end{minted}

\noindent
\revis{in PolaRS. As a result, it requires learning new syntax.}

\section{Related Work}
\label{sec:Related}

\paragraph{\textbf{Optimizing Pandas}} Most previous work on optimizing
dataframe libraries focuses on optimizing \code{pandas}, mainly through the use
of parallel and distributed execution. Systems like \code{modin} \cite{modin},
\code{dask} \cite{dask}, Koalas \cite{koalas_web}, PolaRS \cite{polars_web},
Ponder \cite{ponder_web}, PolyFrame \cite{polyframe} and Magpie \cite{magpie}
are all essentially custom versions of \code{pandas} (some are full rewrites,
while others implement the \code{pandas} API over some underlying system).
Similarly, even techniques like BELE \cite{best_effort_lazy}, whose optimization
targets the \code{pandas}-Python interface, are limited \textit{within} the
library's boundaries, which weakens the view and control of the surrounding
code. In contrast to all these techniques, \system{} is the first system to use
dynamic rewriting at the \code{pandas}-Python interface, and it does so
\textit{externally}, which lets it views all user's code compared to just
library code and can modify any part of it. This is the main conceptual
difference, but there are also other practical drawbacks as we outlined
earlier, mainly arising from the fact that these systems do not focus on
single-machine, ad-hoc, diverse use cases.

\revis{There is also previous work on optimizing \code{pandas} code using static
analysis \cite{opt_ds_static_anal}, which also utilizes library-specific
knowledge. \system{} is different in that it is a dynamic rewriter.
Theoretically, the same rules we use could be applied with static analysis, but
static analysis in Python is quite inaccurate and it does not fit the
read-eval-print-loop (REPL) workflow of EDA workloads.}

\paragraph{\textbf{Pandas for Interactive Settings}} A slightly different and
interesting line of work focuses on optimizing dataframe queries for interactive
workloads \cite{pandas_interactive, lux}. Some of their optimizations include
displaying partial results, reordering operations and performing computation
during \emph{think-time}, i.e., when the user is inspecting results. We also
recognize the importance of interactive workloads, which include the EDA,
single-machine, ad-hoc workloads we focus on in this paper, but we are taking a
different path in optimizing them. We use rewriting at the interface boundary,
which is fundamentally different from the techniques used in this previous work.

\paragraph{\textbf{Rewrite systems in compilers}} Program rewriting is prevalent in compilers. Production-level compilers use peephole optimizers to perform local rewrites. LLVM~\cite{llvm} uses InstCombine \cite{llvm_instcombine} and VectorCombine \cite{llvm_vectorcombine} to perform IR rewrites on scalars and vectors respectively. Further, there have been many works such as Alive~\cite{alive}, Alive2~\cite{alive2}, Souper~\cite{souper} that try to prove or automatically find such rewrites inside traditional compilers. TASO~\cite{taso} and PET~\cite{pet} have looked into how rewrites can be used to optimize tensor computations in tensor compilers. Domain specific languages such as Halide~\cite{halide} include extensive rewrite engines to perform optimizations~\cite{halide-rewrites}. Even complicated optimization passes such as dataflow optimizations~\cite{multiple-rewrites} and vectorization~\cite{vegen} can be expressed as a series of rewrites. In fact, the compiler infrastructure MLIR~\cite{mlir} is rooted on the premise of rewriting to express complex IR transformations. \system{} takes inspiration from these systems that mainly perform static program rewrites and performs rewrites for \code{pandas} implemented in the dynamically-typed Python language.

\paragraph{\textbf{Dynamic Optimization}} There has been a large body of work
that optimizes programs at runtime. Just-in-time (JIT) compilation is one common
technique applied to interpreted languages like Javascript (TraceMonkey
\cite{trace-monkey}, V8 \cite{v8-engine}) and recently Python
\cite{python-spec-adapt-interp}, but also non-native languages like Java
(HotSpot \cite{java-hotspot}). However, all these methods optimize the host
language, focusing on low-level optimizations and not the host-library
combination. On the other hand, our technique can perform higher-level (and
higher-impact) improvements because it understands the semantics of both the
host language and the library.

\section{Conclusion}

In this paper, we identified program rewriting as a lightweight technique for
optimizing ad-hoc, single-machine EDA workloads. We implemented \system{}, the
first source-to-source, dynamic rewriter for Python, system which rewrites
\code{pandas} code automatically and transparently, while simultaneously
addressing the requirements and constraints of condition-checking.

We experimentally showed that \system{} was able to achieve significant speedups
(up to 57$\times$ for individual cells and 3.5$\times$ for whole notebooks),
both compared to \code{pandas} and \code{modin}, in \emph{real-world}, randomly
sampled notebooks. At the same time, \system{} incurs minimal runtime and memory
overheads whether it succeeds or not.

\begin{acks}
We would like to thank Marc Canby, Stratos Vamvourellis, Edward Gan and the anonymous reviewers for insightful comments and suggestions. This work was supported by the AWS Cloud Credit for Research and the Open Philanthropy project.
\end{acks}

\appendix

\section{Extended Results}

\begin{figure*}[!h]
  \centering
  \includegraphics[width=\textwidth]{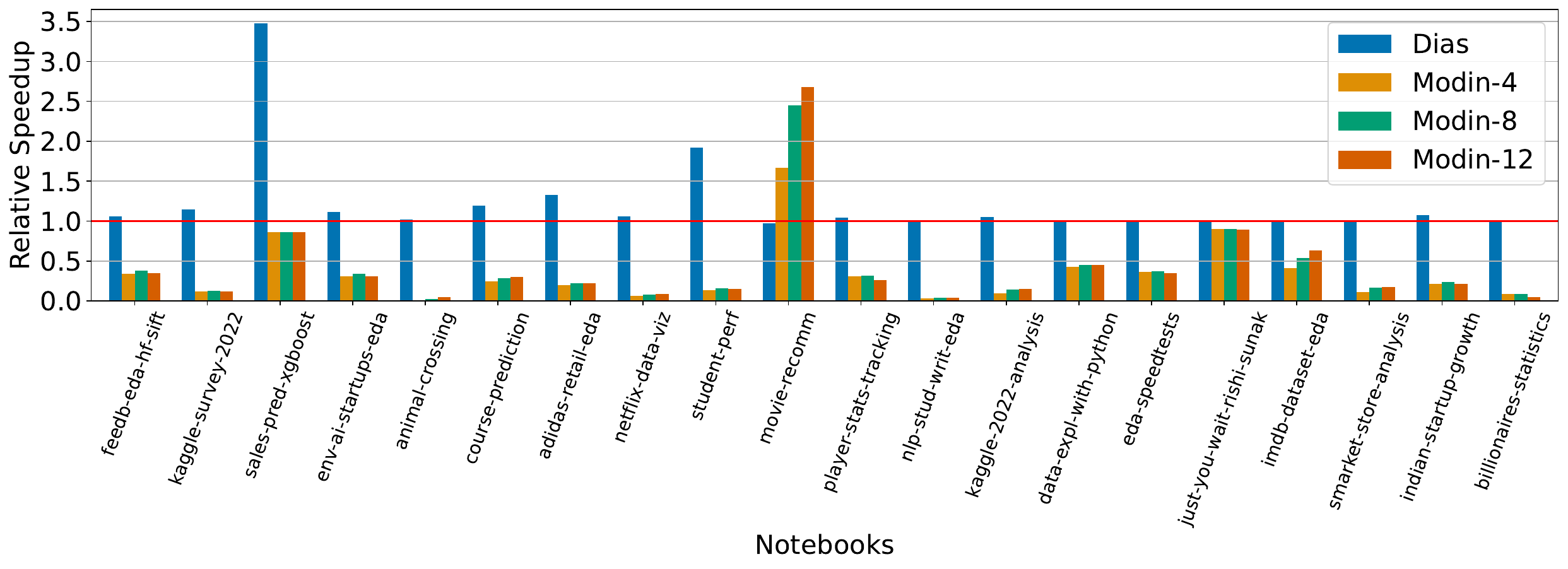}
  \caption{Corresponding to Figure~\ref{fig:modin_nb_level}. The conclusions
  are similar as \code{modin} slows down all the ten new notebooks.}
  \label{fig:ext:modin_nb_level}
\end{figure*}

\begin{figure}[!h]
  \centering
  \includegraphics[width=0.7\columnwidth]{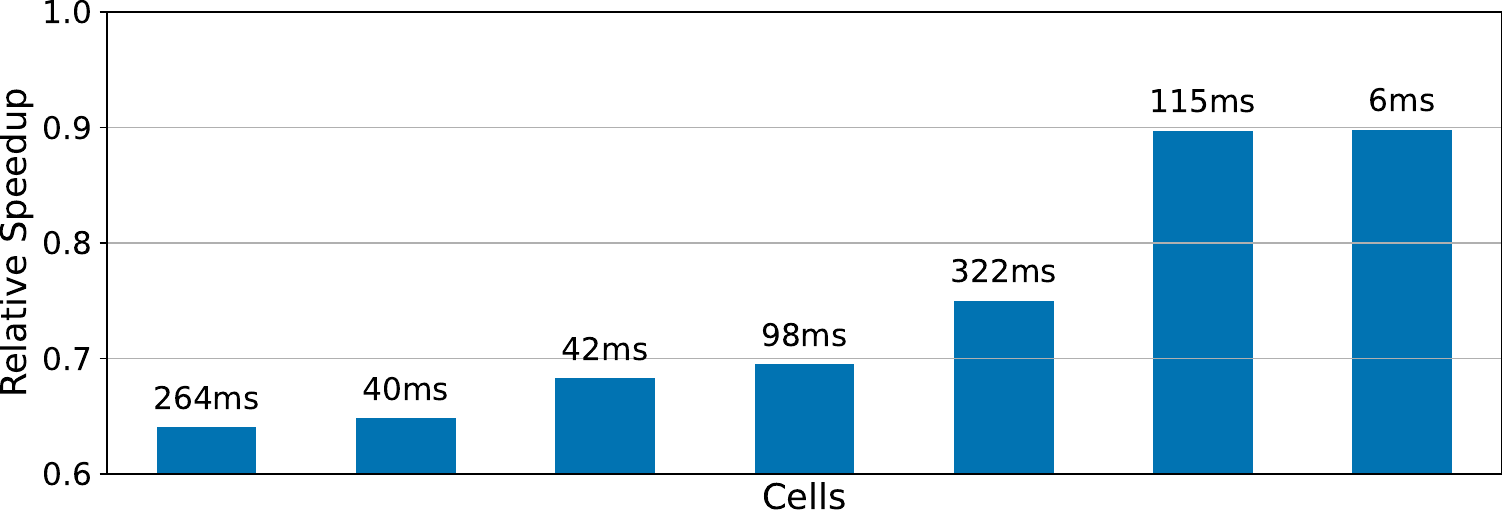}
  \caption{Corresponding to Figure~\ref{fig:cells-slowdowns}. The conclusions
  are the same. The slowdowns are within interactive latencies (i.e., less than
  300ms).}
  \label{fig:ext:cells-slowdowns}
\end{figure}

\begin{figure}[!h]
  \centering
  \includegraphics[width=0.7\columnwidth]{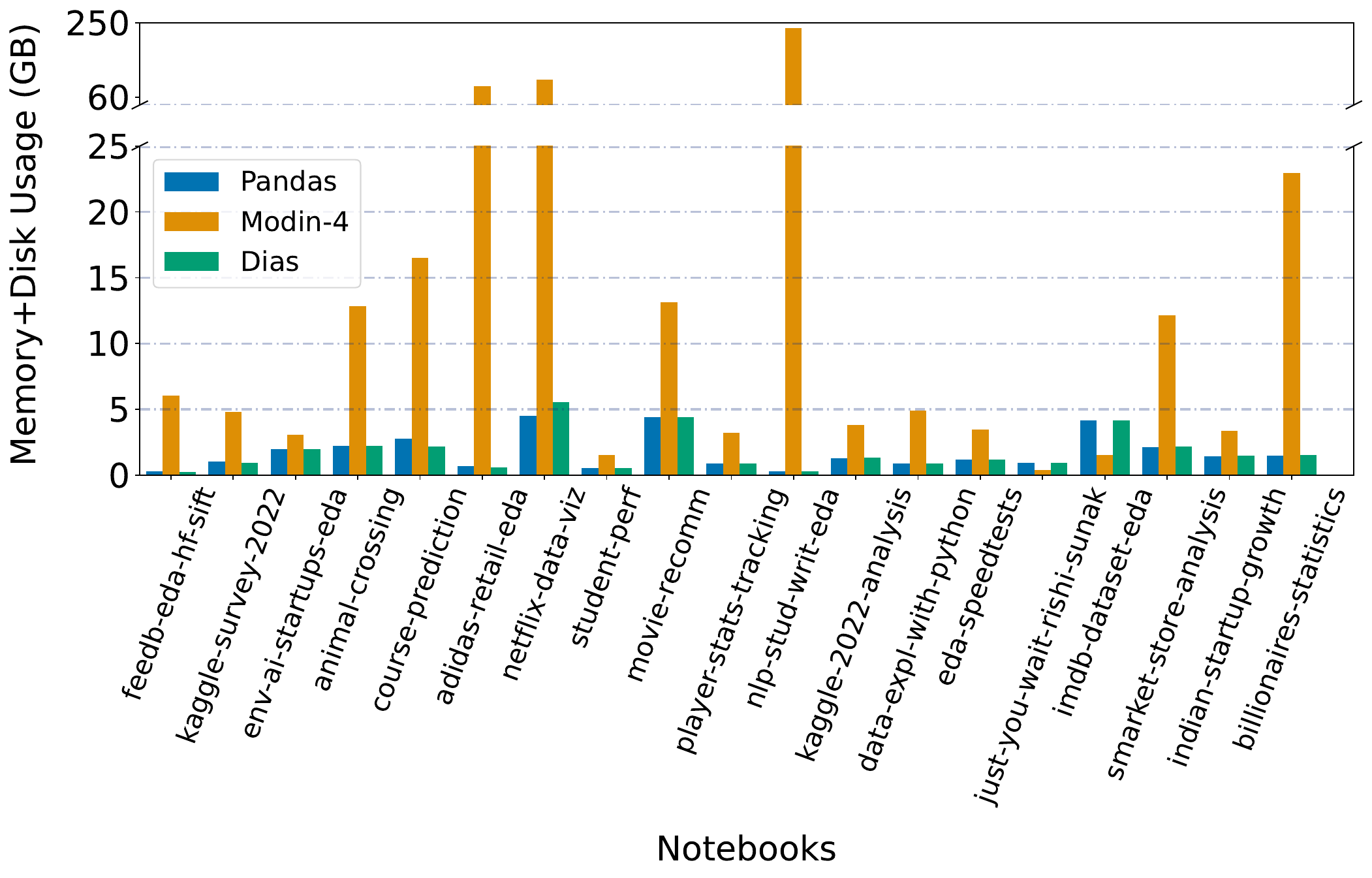}
  \caption{Corresponding to Figure~\ref{fig:modin_nb_mem}. When we include all
  20 notebooks, we see even more aggressive memory+disk usage from
  \code{modin}. \system{} and \code{pandas} remain on the same scales.}
  \label{fig:ext:modin_nb_mem}
\end{figure}

\begin{figure}[!h]
    \centering
    \includegraphics[width=0.8\columnwidth]{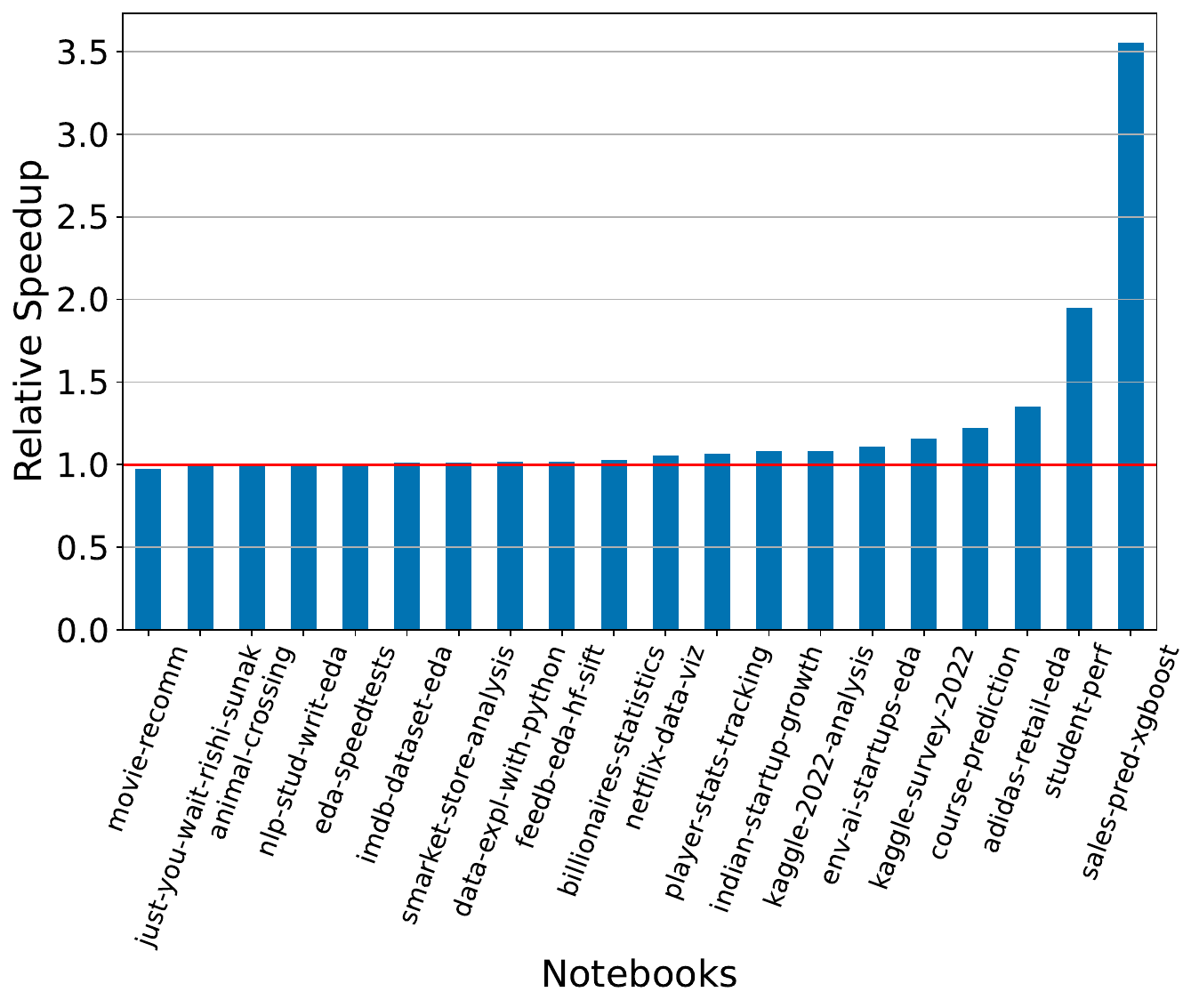}
    \caption{Relative speedups on whole notebooks. We see the same speedups as
    in Figure~\ref{fig:nb_level} with no extra substantial slowdowns when
    considering all 20 notebooks.}
    \label{fig:ext:nb_level}
\end{figure}

\begin{figure*}[!h]
    \centering
    \includegraphics[width=\textwidth]{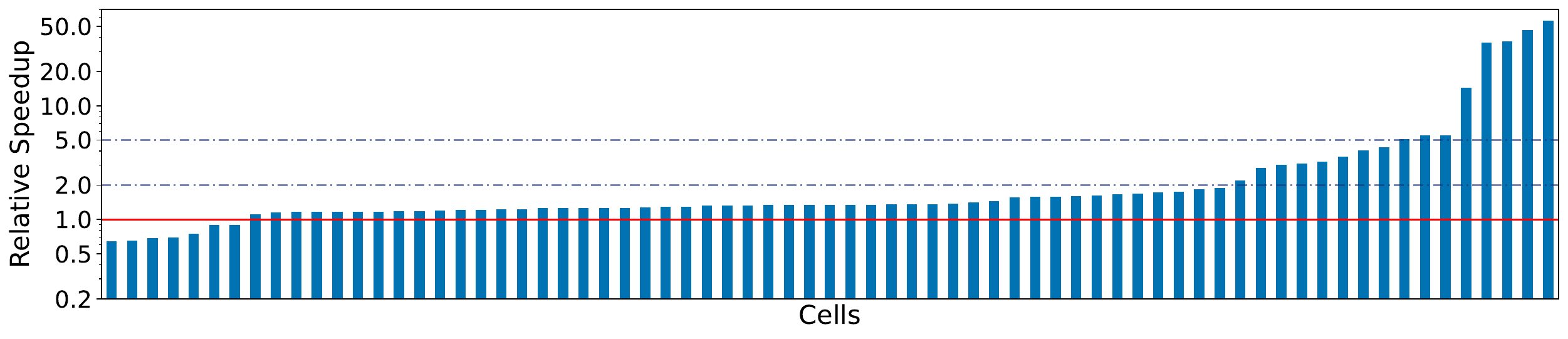}
    \caption{Cell-level relative speedups (excluding cells that originally ran
    for less than 50ms and also all the cells that got a speedup or slowdown
    within the 0.1$\times$ range) for all 20 notebooks. Still, \system{} provides
    significant speedups with no substantial slowdowns (see
    Figure~\ref{fig:ext:cells-slowdowns}).}
    \label{fig:ext:cell_level}
\end{figure*}

In Section~\ref{sec:Evaluation} we focused only on ten out of the twenty random
notebooks we picked (see Section~\ref{sub-sec:exp-setup}). Here, we include
results for all twenty notebooks.

\paragraph{\textbf{Per-Cell Speedups}} Figure~\ref{fig:ext:cell_level} shows the
cell-level speedups, corresponding to Figure~\ref{fig:cell_level}. The plots
look almost identical, and this is because \system{} does not decelerate
notebooks it does not rewrite. Thus, since this plot includes only slowdowns or
speedups that are outside the 0.1$\times$ range, there is hardly any discernible
difference. Similar observations are derived from
Figure~\ref{fig:ext:cells-slowdowns}, where the slowdowns are still under
interactive latency, i.e., 300ms.

These results further validate our hypothesis in
Section~\ref{sub-sec:vs-pandas}. That is, the slowdowns we observe are the
result of rewriting, independent of who performs it (in this case, \system{}).

When we include all twenty notebooks, the geometric mean speedup is 1.18$\times$
and the maximum slowdown is 1.56$\times$.

\paragraph{\textbf{Per-Notebook Speedups}} In Figure~\ref{fig:ext:nb_level} we
show the notebook-level speedups. This figure corresponds to
Figure~\ref{fig:nb_level}. As we mentioned, \system{} does not rewrite code in
the the ten new notebooks, so we do not see any additional speedup. However, it
remains that the slowdowns, when \system{} does not succeed, are minimal. The
geometric mean speedup is now 1.16$\times$ while the maximum slowdown is still
$1.03\times$.

\paragraph{\textbf{Comparison with Modin}} In
Figure~\ref{fig:ext:modin_nb_level} \footnote{which corresponds to
Figure~\ref{fig:modin_nb_level}.}, our conclusions are again unaltered.
\code{modin} slows down all the ten new notebooks and it rarely scales with the
number of cores. The geometric mean and maximum speedup remain the same.

In Figure~\ref{fig:ext:modin_nb_mem}\footnote{which corresponds to
Figure~\ref{fig:modin_nb_mem}.} we show the memory consumption when we consider
all twenty notebooks. The results are not significantly different for
\code{pandas} and \system{}. However, \code{modin}'s memory consumption becomes
even more aggressive. We see that for one notebook, \code{modin} consumes almost
250GB when \code{pandas} and \system{} consume less than 5GB.

\bibliographystyle{ACM-Reference-Format}
\bibliography{sample, rewrite, jit}

\end{document}